\documentclass[12pt]{article}
\usepackage{amsmath,amsfonts,amssymb}
\usepackage{graphicx,xcolor} 
\usepackage{enumitem}
\usepackage{booktabs}
\usepackage{longtable}
\usepackage{siunitx}
\usepackage[justification=justified, singlelinecheck=false]{caption}
\usepackage[colorlinks=true,linkcolor=blue,urlcolor=blue,citecolor=magenta]{hyperref}

\title{Superradiance and Quasinormal Modes of Massive Scalar Fields around Kerr Black Holes in Einstein-Maxwell-Dilaton-Axion Theory\\ with Perfect Fluid Dark Matter}

\author{Teparksorn Pengpan\thanks{{\it Present Address:} Bangkok, Thailand (teparksorn.pengpan@gmail.com)} \\
{\small Division of Physical Science, Faculty of Science, Prince of Songkla University} \\{\small Hat Yai 90110, Thailand}}

\date{}

\begin{document}

\maketitle

\begin{abstract}
We investigate the dynamics of massive scalar fields around Kerr black holes in the Einstein–Maxwell–dilaton–axion (EMDA) theory, incorporating the effects of perfect fluid dark matter (PFDM), characterized by the dilaton parameter \( r_2 \) and the PFDM parameter \( \lambda \). These parameters modify both the ergosphere and the effective potential experienced by the scalar field. Using the asymptotic matching method, we compute superradiant amplification factors, while quasinormal mode (QNM) frequencies are obtained via the asymptotic iteration method (AIM). Our results reveal contrasting effects: increasing \( r_2 \) enhances superradiance and leads to higher QNM frequencies with greater damping, whereas increasing \( \lambda \) suppresses superradiance and reduces QNM frequencies with weaker damping. In combined scenarios, the influence of \( \lambda \) is found to be dominant. These findings extend the understanding of Kerr black holes in EMDA backgrounds and highlight the stabilizing role of PFDM in such systems. 
\end{abstract}

\section{Introduction}
Black holes serve as powerful laboratories for exploring fundamental physics, bridging general relativity with particle physics and cosmology~\cite{Arvanitaki2010,Arvanitaki2011,Rubio2018,Brito2020,Brito2025}. Among their remarkable properties, superradiance is a process in which bosonic fields extract rotational energy from spinning black holes, amplifying outgoing waves when the wave frequency satisfies \( \omega < m \Omega_h \), where \( \Omega_h \) is the horizon angular velocity, defined as the angular frequency associated with the dragging of inertial frames~\cite{Zel'Dovich1971,Zel'Dovich1972,Press1972,Starobinski1973,Press1973,Teukolsky1974,Ford1975}. This phenomenon, akin to the Penrose process for particles~\cite{Penrose1969,Penrose1971a,Penrose1971b}, constrains ultralight particles such as axions~\cite{Dorlis2025a,Dorlis2025b} and provides insights into black hole stability, potential gravitational-wave signatures, and the possible formation of scalar clouds or “black hole bombs” in the presence of instabilities~\cite{detweiler1980black,Zouros1979,Cardoso2005,Endlich2017,Siemonsen2020}. In alternative spacetimes such as regular black holes—nonsingular solutions that circumvent issues including the weak cosmic censorship conjecture and the firewall paradox—superradiance offers a powerful probe of spacetime structure. It can reveal differences between regular and singular geometries through variations in energy-extraction efficiency and instability growth rates, with potential observational signatures in black hole spin-down, mass–spin parameter gaps, and gravitational-wave emissions from boson annihilations or level transitions~\cite{Yang2023}.

Quasinormal modes (QNMs) characterize the damped oscillations during the ringdown phase of black hole perturbations, encoding the black hole’s mass \(M\), spin \(a\), and electric charge \(Q\). They serve as powerful probes for testing modified gravity theories and assessing environmental effects~\cite{Berti2009,Kokkotas1999,Konoplya2011,Berti2016PRL,Yang2021,Sun2023,Das2024,Wang2024,Toshmatov2025,Tan2025}. 
Recent astrophysical observations, including gravitational-wave detections by LIGO/Virgo/KAGRA~\cite{Abbott2016,Abbott2020} and Event Horizon Telescope images of supermassive black holes~\cite{Akiyama2019,Akiyama2022}, highlight the importance of studying superradiance and QNMs, which offer testable predictions for ultralight bosons, dark matter interactions, and deviations from Kerr geometry, potentially revealing new physics through waveform mismatches or spectral instabilities.

Dark matter, a significant fraction of the universe’s mass-energy, forms halos around galaxies and modifies black hole spacetimes \cite{navarro1997universal,Oliver2018}. For incorporating the dark matter’s gravitational effects into black holes, the perfect fluid dark matter (PFDM) model offers an analytical framework. With the Newman-Janis algorithm it can be extended to rotating black holes~\cite{Li2012,azreg2014newman,Liu2023superradiance}. Within the Einstein-Maxwell-dilaton-axion (EMDA) theory, which includes dilaton and axion fields alongside electromagnetic interactions \cite{Garcia1995,Ganguly2016,Banerjee2020,Senjaya2025}, PFDM enables the study of complex gravitational and matter dynamics \cite{Das2021,Liang2023}. The interplay between PFDM and EMDA parameters is particularly intriguing, as dark matter alters spacetime geometry, potentially affecting the superradiance amplification factor and providing a novel probe for dark matter properties \cite{Endlich2017,Siemonsen2020}.

In this work, we investigate the superradiance of a massive scalar field in a Kerr black hole surrounded by perfect fluid dark matter (PFDM) within the Einstein–Maxwell–dilaton–axion (EMDA) framework. Using a Mathematica-based symbolic–numerical pipeline, we derive the separated equations, compute the amplification factor \(Z_{lm}\) via asymptotic matching~\cite{detweiler1980black,Furuhashi2004,Bao2022,Bao2023}, and evaluate quasinormal modes (QNMs) with the asymptotic iteration method using a two-pass update of the angular separation constant (from \texttt{SpheroidalEigenvalue}), benchmarked against continued-fraction results in the Kerr limit~\cite{Leaver1985,Konoplya2006,Dolan2007,Pan2006}. We interpret trends via the effective potential in tortoise coordinates. The PFDM parameter \(\lambda\) and dilaton parameter \(r_2\), constrained by observations~\cite{Akiyama2022}, modify the spacetime and imprint on both \(Z_{lm}\) and the QNM spectrum. Throughout, we adopt geometrized units \(G=c=1\) and set \(M=1\), retaining explicit factors of \(M\) in analytic expressions to highlight scaling. 

\section{Methodology}

\subsection{EMDA Theory with PFDM}
\label{sec:EMDA_Theory_with_PFDM}

The Einstein–Maxwell–dilaton–axion (EMDA) theory describes the coupled dynamics of gravity, electromagnetism, the dilaton \(\varphi\), and the axion \(\chi\), and is given by~\cite{Garcia1995,Ganguly2016,Banerjee2020,Senjaya2025,Wei2013,Flathmann2015}
\begin{equation}
S_{\mathrm{EMDA}} = \frac{1}{16\pi} \int d^4x \,\sqrt{-g}\,\left[ R + \mathcal{L}_{\mathrm{EMDA}}\right],
\end{equation}
where \(g \equiv \det(g_{\mu\nu})\) and \(R\) is the Ricci scalar. The Lagrangian density is
\begin{equation}
\mathcal{L}_{\mathrm{EMDA}} = - 2\, g^{\mu\nu} \partial_\mu \varphi\,\partial_\nu \varphi 
- \frac{1}{2} e^{4\varphi} g^{\mu\nu} \partial_\mu \chi\, \partial_\nu \chi 
- e^{-2\varphi} F_{\mu\nu}F^{\mu\nu} 
- \chi\,F_{\mu\nu} \tilde{F}^{\mu\nu},
\end{equation}
with \(F_{\mu\nu}\) the Maxwell tensor and \(\tilde{F}^{\mu\nu} = \frac{1}{2} \epsilon^{\mu\nu\alpha\beta}F_{\alpha\beta}\) its dual. The massless axion \(\chi\), dual to the three-form field strength of the Kalb–Ramond tensor~\cite{Kalb1974}, couples to electromagnetism via the last term and can contribute to black hole rotational energy~\cite{Ganguly2016,Banerjee2020}. The dilaton \(\varphi\), ubiquitous in string theory, controls the effective coupling \(g_s \sim e^{\varphi}\) and couples non-minimally to the Maxwell sector~\cite{Herdeiro2025}, producing black hole scalar “hair.”

To include perfect fluid dark matter (PFDM)~\cite{Li2012,Liu2023superradiance}, we add its stress-energy tensor \(T_{\mathrm{PFDM}}^{\mu\nu}\) to the Einstein equations without direct coupling to the EMDA fields:
\begin{equation}
G_{\mu\nu} \equiv R_{\mu\nu} - \frac{1}{2}g_{\mu\nu}R = 8\pi\left(T_{\mathrm{EMDA}}^{\mu\nu} + T_{\mathrm{PFDM}}^{\mu\nu}\right),
\end{equation}
where \(T_{\mathrm{EMDA}}^{\mu\nu}\) is obtained from \(\mathcal{L}_{\mathrm{EMDA}}\) and \(T^\mu_{\ \nu}{}_{\mathrm{PFDM}} = \mathrm{diag}(-\rho_{\mathrm{DM}}, 0, 0, 0)\) with the total energy density  \(\rho_{\mathrm{DM}} = \lambda/(8\pi r^3)\). Here, the parameter \(\lambda\) is proportional to the PFDM mass density~\cite{Qiao2023}.  The PFDM parameter \(\lambda \geq 0\) satisfies the weak energy condition and yields a logarithmic modification \(-\lambda r \ln(r/\lambda)\) to the metric function \(\Delta\). For \(\lambda/M \lesssim 0.15\) (\(2\sigma\))~\cite{Vagnozzi2023}, consistent with astrophysical bounds, the PFDM backreaction is linear and does not induce significant nonlinear effects. The EMDA field equations for \(\varphi\), \(\chi\), and \(F_{\mu\nu}\) remain unchanged.  

Spherically symmetric EMDA black hole solutions can be extended to axisymmetry via the Newman–Janis algorithm, yielding the Kerr–Sen metric~\cite{Sen1992}.  A stationary, axisymmetric EMDA–PFDM black hole in Boyer–Lindquist coordinates takes the form
\begin{equation}
\label{Eq:EMDA-PFDM_metric}
\begin{aligned}
ds^2 &= -\frac{\Delta_D - a^2\sin^2\theta}{\rho^2}\,dt^2 + \frac{\rho^2}{\Delta_D}\,dr^2 + \rho^2\,d\theta^2 \\
&\quad + \frac{2a\sin^2\theta}{\rho^2}\left[ r(r+2r_2) + a^2 - \Delta_D \right] dt\,d\phi \\
&\quad + \frac{\sin^2\theta}{\rho^2} \left[ \left(r(r+2r_2) + a^2\right)^2 - \Delta_D a^2 \sin^2\theta \right] d\phi^2,
\end{aligned}
\end{equation}
with
\begin{equation}
\label{Def:rho2_DeltaD}
\rho^2 = r(r+2r_2) + a^2\cos^2\theta,\quad
\Delta_D(r) = r(r+2r_2) + a^2 - 2Mr - \lambda r \ln\left(\frac{r}{\lambda}\right).
\end{equation}
For \(\lambda=0\), Eq.~\eqref{Eq:EMDA-PFDM_metric} reduces to the \(b=0\) EMDA solution of García–Gal’tsov–Kechkin~\cite{Garcia1995} under the mappings \(\rho^2 \mapsto \Delta,\ \Delta_D \mapsto \Sigma,\ r_2 \mapsto -\beta,\ M-r_2 \mapsto m\). The dilaton parameter \(r_2 = (Q e^{\phi_0})^2/(2M)\) depends on the dilaton field \(\phi_0\), mass \(M\), and charge \(Q\). The PFDM term preserves the Kerr–Sen structure while adding the logarithmic backreaction.            

Using Mathematica's symbolic computation capabilities, we verify that the Einstein tensor constructed from the EMDA-PFDM metric (Eq. \eqref{Eq:EMDA-PFDM_metric}) satisfies the condition \(\nabla_\mu G^\mu_\nu =0\). This result confirms the conservation of the corresponding stress-energy tensor:
\begin{equation} 
\label{Eq:zero-divergence_stress-energy_tensor}
\nabla_\mu T^\mu_{\nu,\text{PFDM}}=0.
\end{equation}
We refrain from presenting the complete explicit form of the stress-energy tensor \(T^\mu_{\nu,\text{PFDM}}\) here, as its components are rather lengthy and complicated. However, in the limiting case \(r_2 \to 0\) and \(a \to 0\), it reduces to the much simpler diagonal expression
\begin{equation} 
\label{Eq:stress-energy_tensor}
T^\mu_{\ \nu,\text{PFDM}} =
\begin{pmatrix}
\frac{\lambda}{8\pi r^3} & 0 & 0 & 0 \\
0 & \frac{\lambda}{8\pi r^3} & 0 & 0 \\
0 & 0 & -\frac{\lambda}{16\pi r^3} & 0 \\
0 & 0 & 0 & -\frac{\lambda}{16\pi r^3}
\end{pmatrix}.
\end{equation}

\begin{figure}[h]
\centering 
\includegraphics[width=\textwidth]{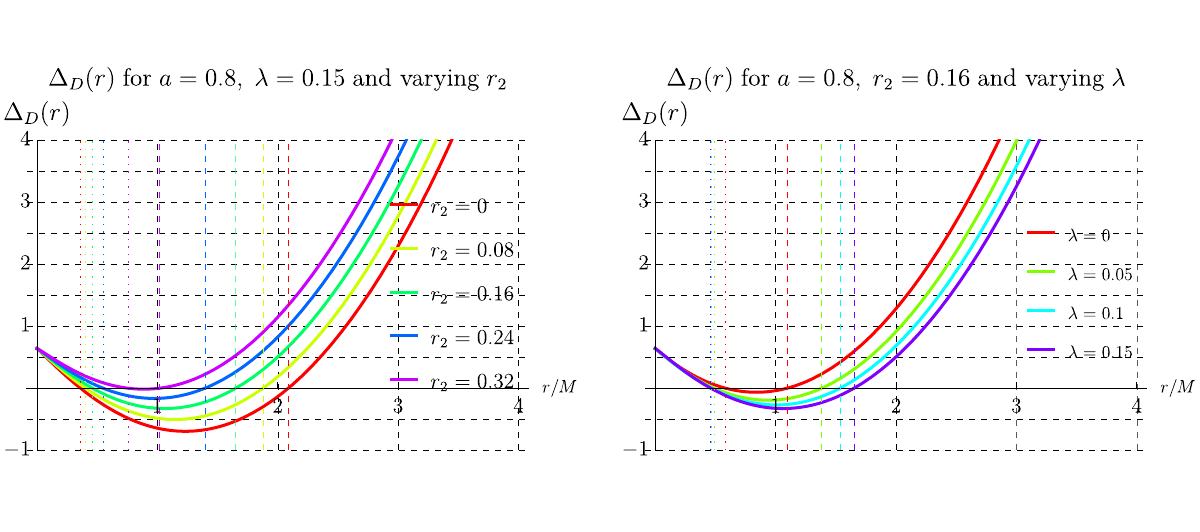}
\caption{Effect of \(r_2\) (left) and \(\lambda\) (right) on \(\Delta_D(r)\) for a Kerr black hole with \(M=1\) and \(a=0.8\).}
\label{fig:deltaplots}
\end{figure}

The event horizon ($  r_h  $) and Cauchy horizon ($  r_c  $) are the real roots of $  \Delta_D(r)=0  $. Figure~\ref{fig:deltaplots} shows $  \Delta_D(r)  $ for a fixed spin parameter $  a  $ and varying $  r_2  $ and $  \lambda  $. The requirement that $  r_h  $ and $  r_c  $ be real roots of $  \Delta_D(r)=0  $ constrains the allowed ranges of $  r_2  $ and $  \lambda  $ depending on the black hole spin $  a  $. For example, at the uppermost astrophysical bound $  \lambda/M=0.15  $ (2$  \sigma  $)~\cite{Vagnozzi2023}, $  r_2  $ must be less than 0.32 for a rapidly rotating black hole with $  a=0.8  $ and less than 0.16 for a near-extremal black hole with $  a=0.995  $. For fixed $  \lambda  $, increasing $  r_2  $ decreases $  r_h  $ relative to the infinite-redshift surface $  r_{\mathrm{rs}}  $ (approximately determined by solving $  g_{tt}=0  $), thereby enlarging the ergosphere. For fixed $  r_2  $, increasing $  \lambda  $ increases $  r_h  $ relative to $  r_{\mathrm{rs}}  $, shrinking the ergosphere. Numerical values are given in Table~\ref{Ergosphere_varying_r2_lambda} (Sec.~\ref{subSec:Energy_Extraction}).

\subsection{Scalar Field Dynamics}

We consider a massive scalar field \( \Phi \) of mass \( \mu \) propagating in the Kerr black hole spacetime with perfect fluid dark matter (PFDM) in the Einstein–Maxwell–dilaton–axion (EMDA) framework. The dynamics of \( \Phi \) are governed by the Klein–Gordon equation
\begin{equation}
\label{Eq:Klein-Gordon}
\left( \nabla^\mu \nabla_\mu - \mu^2 \right) \Phi  
= \frac{1}{\sqrt{-g}} \partial_\mu \left( \sqrt{-g} \, g^{\mu\nu} \partial_\nu \Phi \right) - \mu^2 \Phi = 0.
\end{equation}

In the background metric \eqref{Eq:EMDA-PFDM_metric}, we adopt the separable ansatz~\cite{detweiler1980black,Deruelle1974,Damou1976}
\begin{equation*}
\Phi(t, r, \theta, \phi) = e^{-i \omega t} e^{i m \phi} S(\theta) R(r),
\end{equation*}
which allows Eq.~\eqref{Eq:Klein-Gordon} to separate into angular and radial parts:
\begin{equation}
\label{Eq:angular_part}
\frac{1}{\sin \theta} \frac{d}{d\theta} \left( \sin \theta \frac{dS}{d\theta} \right) + \left[ \Lambda_{lm} - a^2 (\mu^2 - \omega^2) \cos^2 \theta - \frac{m^2}{\sin^2 \theta} \right] S = 0,
\end{equation}
\begin{equation}
\label{Eq:radial_part}
\begin{aligned}
\Delta_D(r) \frac{d}{dr} & \left[ \Delta_D(r) \frac{dR}{dr} \right] + \left[ \left( \omega \left[ r(r + 2 r_2) + a^2 \right] - a m \right)^2 \right. \\
& \left. - \Delta_D(r) \left( \Lambda_{lm} + r(r + 2 r_2) \mu^2 - 2 a m \omega + a^2 \omega^2 \right) \right] R = 0,
\end{aligned}
\end{equation}
where \( \Lambda_{lm} \) is the separation constant~\cite{Teukolsky1973ApJ,Berti2006}.  
The radial equation~\eqref{Eq:radial_part} serves as the basis for the analysis of absorption cross sections, quasi-bound states, and quasinormal modes.

For the angular equation~\eqref{Eq:angular_part}, introducing \( u = \cos\theta \) yields
\begin{equation}
\label{Equ:angular_part}
\frac{d}{du} \left[ (1 - u^2) \frac{dS(u)}{du} \right] + \left[ \Lambda_{lm} - a^2 (\mu^2 - \omega^2) u^2 - \frac{m^2}{1 - u^2} \right] S(u) = 0.
\end{equation}
The asymptotic behavior near the poles \( u \to \pm 1 \) is
\begin{equation}
S(u) \sim 
\begin{cases}
(1 - u)^{|m|/2}, & u \to 1, \\
(1 + u)^{|m|/2}, & u \to -1.
\end{cases}
\end{equation}

For the radial sector, we work in the domain \( r \in [r_h, \infty) \) outside the event horizon. To remove coordinate singularities, we introduce the tortoise coordinate \( r_* \) via
\begin{equation}
dr_* = \frac{r(r + 2 r_2) + a^2}{\Delta_D(r)} \, dr,
\end{equation}
which maps \( r \in [r_h, \infty) \) to \( r_* \in (-\infty, \infty) \). Defining
\begin{equation*}
R(r) = \frac{\Psi(r_*)}{\sqrt{r(r + 2 r_2) + a^2}},
\end{equation*}
the radial equation~\eqref{Eq:radial_part} transforms into a Schrödinger-like form,
\begin{equation}
\label{Eq:Schrödinger-like}
\frac{d^2 \Psi}{dr_*^2} + \left[ \hat{\omega}^2 - V_{\text{eff}}(r) \right] \Psi = 0,
\end{equation}
with the shifted frequency
\begin{equation*}
\hat{\omega} = \omega - \frac{a m}{r(r + 2 r_2) + a^2},
\end{equation*}
and the effective potential
\begin{equation}
\label{Eq:Veff}
\begin{aligned}
V_{\text{eff}}(r) &= \frac{\left[ \Lambda_{lm} + r(r + 2 r_2) \mu^2 + a \omega (-2 m + a \omega) \right] \Delta_D(r)}{\left[ r(r + 2 r_2) + a^2 \right]^2} \\
&\quad + \frac{\Delta_D(r)}{\left[ r(r + 2 r_2) + a^2 \right]^2} \left[ \frac{\Delta_D'(r)}{2} - \frac{\Delta_D(r)(r + r_2)}{r(r + 2 r_2) + a^2} \right] \\
&\quad + \frac{3 \, \Delta_D(r)^2 (r + r_2)^2}{\left[ r(r + 2 r_2) + a^2 \right]^3} - \frac{\Delta_D(r)^2}{\left[ r(r + 2 r_2) + a^2 \right]^2}.
\end{aligned}
\end{equation}
This transformation eliminates the first-derivative term, thereby simplifying the analysis. The effective potential contains:  
(i) a rotational term, modified by \( r_2 \) and reflecting PFDM-induced frame dragging;  
(ii) mass and angular momentum contributions, also dependent on \( r_2 \), with \( \Lambda_{lm} \) encoding spin effects;  
(iii) PFDM-induced corrections via \( \Delta_D(r) \) that depend on \( \lambda \); and  
(iv) geometric terms involving derivatives of \( \Delta_D(r) \), sensitive to surface gravity and \( r_2 \).

\begin{figure}[h]
\centering
\includegraphics[width=\textwidth]{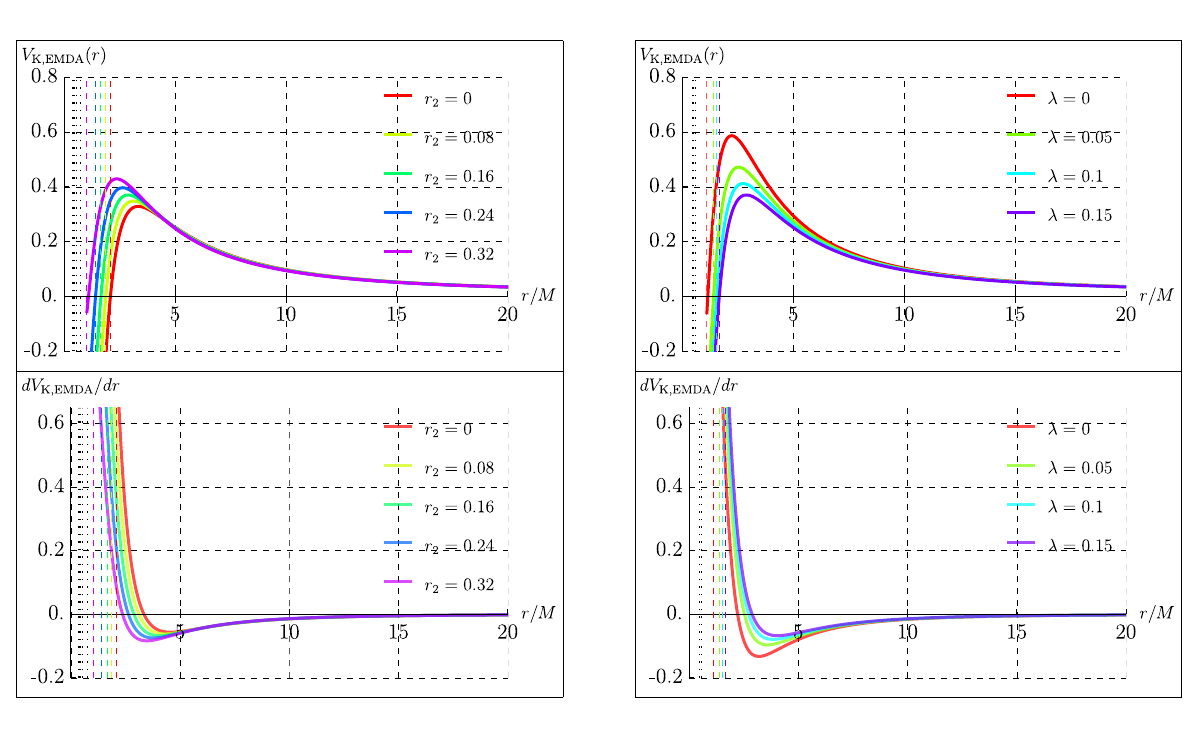}
\caption{
Profiles of the Kerr–EMDA effective potential for a massive scalar field with real frequency. 
(a) Left panel: variation with the dilaton parameter \( r_2 \) at fixed PFDM parameter \( \lambda = 0.15 \). 
(b) Right panel: variation with \( \lambda \) at fixed \( r_2 = 0.16 \). 
The black hole mass and spin are \( M = 1 \) and \( a = 0.8 \), with azimuthal mode \( m = 1 \), scalar mass \( \mu = 0.1 \), and real frequency \( \omega = 0.08 \). 
Changes in \( r_2 \) and \( \lambda \) modify the location and height of the potential barrier, thereby influencing the superradiant scattering process.
}
\label{fig:PotentialPlotsRealOmega}
\end{figure}

\begin{figure}
\centering
\includegraphics[width=\textwidth]{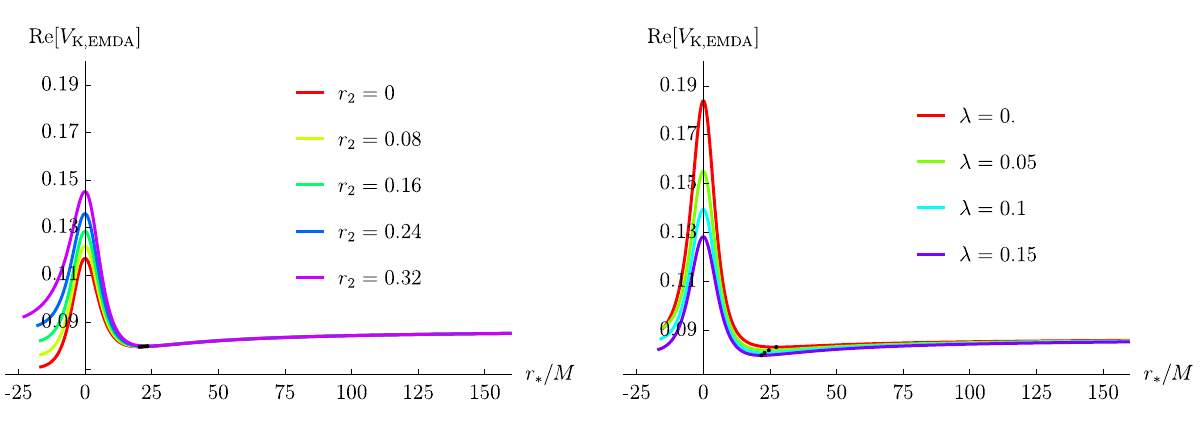}
\includegraphics[width=\textwidth]{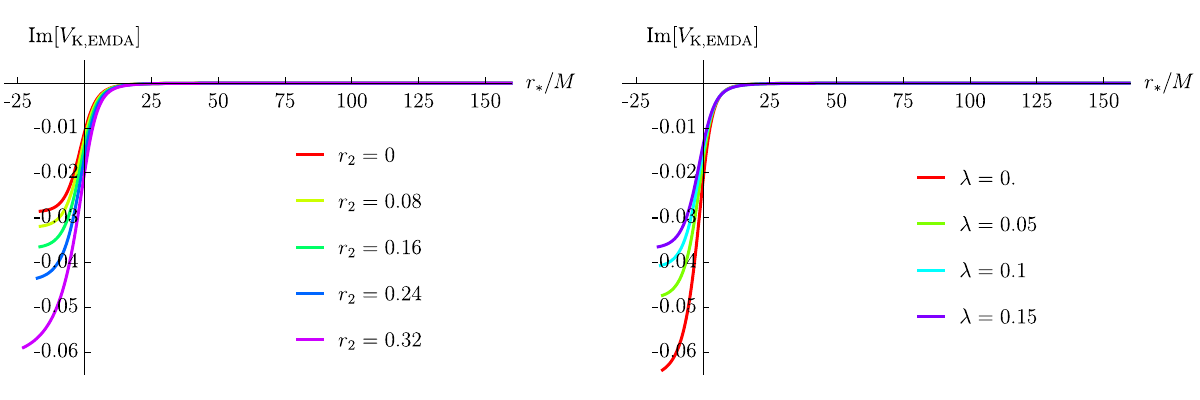}
\includegraphics[width=\textwidth]{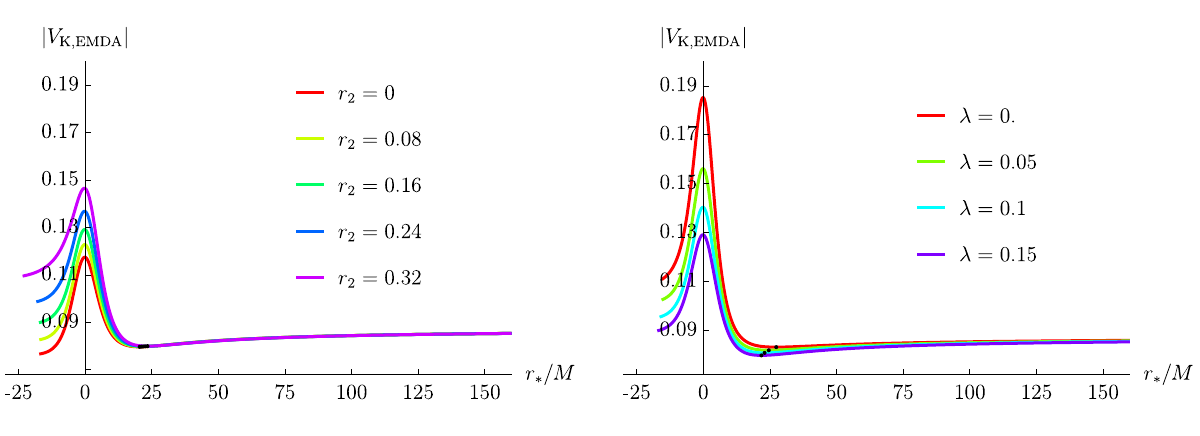}
\caption{
Profiles of the Kerr–EMDA effective potential for a massive scalar field with complex frequency, complementing the real-frequency case shown in Fig.~\ref{fig:PotentialPlotsRealOmega}. 
Real, imaginary, and absolute values of the potential are plotted as functions of the tortoise coordinate \( r_* \) for \( \omega = 0.2 - 0.1 i \). 
(a) Left panels: variation with the dilaton parameter \( r_2 \) at fixed PFDM parameter \( \lambda = 0.15 \). 
(b) Right panels: variation with \( \lambda \) at fixed \( r_2 = 0.16 \). 
Black dots indicate the local minima of the potential profiles. 
The parameters are \( M = 1 \), \( a = 0.8 \), \( m = 1 \), and \( \mu = 0.1 \). 
Variations in \( r_2 \) and \( \lambda \) shift the potential minima and alter the barrier shape, affecting mode stability and amplification.
}
\label{fig:PotentialPlotsComplexOmega}
\end{figure}

Here, $V_{\rm K,EMDA}(r)$ denotes the effective potential in the Kerr--EMDA spacetime, obtained from the Schrödinger-like equation~\eqref{Eq:Schrödinger-like}.  Figure~\ref{fig:PotentialPlotsRealOmega} shows representative profiles of $V_{\rm K,EMDA}$ as a function of the radial coordinate $r$ for massive scalar fields with real-frequency modes. Figure~\ref{fig:PotentialPlotsComplexOmega} displays the real part (top), imaginary part (middle), and absolute value (bottom) of $V_{\rm K,EMDA}$ as a function of the tortoise coordinate $r_*$ for massive scalar fields with complex-frequency modes.

For complex-frequency modes the effective potential is complex-valued. The trapping well relevant for quasi-bound states and superradiant instability is defined exclusively from its real part, since $\operatorname{Re}[V_{\rm K,EMDA}]$ determines the conservative confining structure of the Schrödinger-like equation, while $\operatorname{Im}[V_{\rm K,EMDA}]$ primarily encodes dissipative or instability effects. The absolute value $|V_{\rm K,EMDA}|$ is shown only for visualization of the overall magnitude and is not used to locate the trapping well. Accordingly, the local minimum is identified by the standard conditions
\begin{equation}
\frac{d\,\operatorname{Re}[V_{\rm K,EMDA}]}{dr_*}=0,
\qquad
\frac{d^2\,\operatorname{Re}[V_{\rm K,EMDA}]}{dr_*^2}>0.
\end{equation}
We note that $\operatorname{Im}[V_{\rm K,EMDA}]$ becomes nearly flat for $r_*/M\gtrsim 15$, further supporting the dominance of the real part in determining the location and depth of the trapping well. These potential profiles control the scattering behaviour of scalar waves: the height and location of the barrier govern the transmission and reflection amplitudes that enter the superradiant amplification factor $Z_{lm}$ discussed in the following section.

The asymptotic limits of \( V_{\text{K,EMDA}} \) are
\begin{equation}
\lim_{r \to r_h} V_{\text{K,EMDA}}(r) = \left( \omega - m \Omega_h \right)^2 \equiv k_h^2, 
\quad 
\Omega_h = \frac{a}{r_h (r_h + 2 r_2) + a^2},
\end{equation}
\begin{equation}
\lim_{r \to \infty} V_{\text{K,EMDA}}(r) = \omega^2 - \mu^2 \equiv k_\infty^2.
\end{equation}
Accordingly, the asymptotic solutions of Eq.~\eqref{Eq:Schrödinger-like} take the form
\begin{equation}
\label{Sols:Schrödinger-like}
\Psi(r_*) \sim 
\begin{cases}
\mathcal{T} \, e^{-i k_h r_*}, & r_* \to -\infty \quad (r \to r_h), \\
\mathcal{I} \, e^{-i k_\infty r_*} + \mathcal{R} \, e^{i k_\infty r_*}, & r_* \to \infty \quad (r \to \infty),
\end{cases}
\end{equation}
where \( \mathcal{I} \), \( \mathcal{R} \), and \( \mathcal{T} \) denote the incident, reflected, and transmitted amplitudes, respectively.  
Equating the Wronskian at the horizon and at infinity yields
\begin{equation}
\label{Eq:reflection_transmission}
|\mathcal{R}|^2 = |\mathcal{I}|^2 - \frac{k_h}{k_\infty} \, |\mathcal{T}|^2.
\end{equation}
Superradiance occurs for massive scalar fields when
\begin{equation*}
\mu < \omega < m \Omega_h,
\end{equation*}
in which case \( |\mathcal{R}|^2 / |\mathcal{I}|^2 > 1 \).  
Here, \( \Omega_h \) decreases with increasing \( \lambda \) due to the outward shift of the event horizon (Fig.~\ref{fig:deltaplots}), thereby narrowing the frequency range for superradiance. The amplification factor is defined as
\begin{equation}
\label{Def:amplification_factor}
Z_{lm} = \frac{|\mathcal{R}|^2}{|\mathcal{I}|^2} - 1,
\end{equation}
which we compute using the asymptotic matching method~\cite{detweiler1980black,Furuhashi2004,Bao2022,Bao2023,Li2023}.

\section{Superradiance: Asymptotic Matching Method}
\label{Sec:Superradiance-Asymptotic_Matching_Method}
The asymptotic matching method~\cite{detweiler1980black} is employed to solve the radial equation~\eqref{Eq:radial_part} in two distinct regimes: (i) the near-horizon region, \( r - r_h \ll \omega^{-1} \), and (ii) the far region, \( r \gg M \). To evaluate the amplification factor \( Z_{lm} \), we match these solutions in an overlap region satisfying \( M \ll r - r_h \ll \omega^{-1} \) under the low-frequency assumption \( \mu M < \omega M \ll 1 \) and for spin values
\( a \leq M\).

\subsection{Near-Horizon Solution}
As shown in Fig.~\ref{fig:deltaplots}, the function \( \Delta_D(r) \) has simple roots at the event horizon \( r_h \) and the Cauchy horizon \( r_c \). In the vicinity of \( r_h \), we may approximate
\begin{equation}
\label{Def:Approx_DeltaD}
\Delta_D(r) \approx (r - r_h)(r - r_c),
\end{equation}
which is valid because higher-order terms in \( r - r_h \) are negligible in the near-horizon expansion. Introducing the dimensionless variable
\begin{equation}
x = \frac{r - r_h}{r_h - r_c},
\end{equation}
the radial equation~\eqref{Eq:radial_part} reduces to
\begin{equation}
\label{Eq:near-horizon}
x^2 (1 + x)^2 R_{lm}''(x) + x (1 + x) (1 + 2x) R_{lm}'(x) + \left[ Q^2 - x (1 + x) l'(l' + 1) \right] R_{lm}(x) = 0,
\end{equation}
where
\begin{equation}
Q = \frac{\omega \left[ r_h (r_h + 2 r_2) + a^2 \right] - a m}{r_h - r_c}, \quad l' = l + \epsilon,
\end{equation}
and \( \epsilon \) accounts for small perturbative corrections due to the slow-rotation and small-\( \mu M \) approximations. The ingoing solution at the event horizon can be written in terms of the Gauss hypergeometric function:
\begin{equation}
R_{lm}(x) \sim   \left( \frac{1 + x}{x} \right)^{i Q} \, _2F_1(-l',\, l' + 1,\, 1 - 2i Q,\, -x),
\end{equation}
whose large-\( x \) asymptotic form reads
\begin{equation}
\label{Sols:near-horizon_large-x}
R_{lm}(x) \sim   x^{l'} \frac{\Gamma(1 - 2i Q) \Gamma(2 l' + 1)}{\Gamma(l' + 1) \Gamma(l' + 1 - 2i Q)}
+  x^{-1 - l'} \frac{\Gamma(1 - 2i Q) \Gamma(-2 l' - 1)}{\Gamma(-l' - 2i Q) \Gamma(-l')}.
\end{equation}

\subsection{Far-Region Solution}

In the far region \( r \gg M \), the leading terms in \(\Delta_D(r)\) scale as \(r^2\), while the PFDM logarithmic term \(-\lambda r \ln(r/\lambda)\) is negligible for small \(\lambda\). Approximating Eq.~\eqref{Def:Approx_DeltaD} and expanding in the limit \(r \to \infty\), the radial equation~\eqref{Eq:radial_part} reduces to
\begin{equation}
\label{Eq:Far_radial_approx}
\Psi''(r) + \left[ -\kappa^2 + \frac{2\kappa \nu}{r} - \frac{l'(l'+1)}{r^2} \right] \Psi(r) = 0,
\end{equation}
where \(\Psi(r) = r R(r)\), \(\kappa = \sqrt{\mu^2 - \omega^2}\),
\begin{equation}
2\kappa\nu = M\left( 2\omega^2 - \mu^2 \right)\,\alpha_1, \quad
l'(l'+1) = \Lambda_{lm} + \alpha_2\,\mu^2 + \alpha_3\,\omega^2,
\end{equation}
and the geometric coefficients \(\alpha_1, \alpha_2, \alpha_3\) are
\begin{equation}
\label{eqs:geometric_coefficients}
\begin{aligned}
\alpha_1 &= r_h + r_c + 2r_2, \\
\alpha_2 &= r_h^2 + r_c r_h + r_c^2 + 2r_2 (r_h + r_c), \\
\alpha_3 &= -3(r_h^2 + r_c^2) - 4(r_2^2 + r_h r_c) - 8r_2 (r_h + r_c).
\end{aligned}
\end{equation}
Equation~\eqref{Eq:Far_radial_approx} is of Coulomb-type, with a decaying solution at infinity given by
\begin{equation}
\Psi(r) \sim 2^{l'+1} e^{-\kappa r} (\kappa r)^{l'+1} \,
U(l' + 1 - \nu,\, 2 l' + 2,\, 2\kappa r),
\end{equation}
where \(U\) is the confluent hypergeometric function of the second kind.  
For small \(r\), this behaves as
\begin{equation}
\label{Sols:far-region_small-r}
R(r) \sim \frac{2^{l'} \kappa^{l'} \Gamma(2l'+1)}{\Gamma(-l' - \nu)}\, r^{l'}
+ \frac{2^{-1-l'} \kappa^{-1-l'} \Gamma(2l'+1)}{\Gamma(l' - \nu + 1)}\, r^{-l'-1}.
\end{equation}

\subsection{Matching Leading-Order Terms}
\label{Matching_Leading-Order_Terms}
In the overlap region, the near-horizon solution~\eqref{Sols:near-horizon_large-x} takes the form
\begin{equation}
R(r) \sim A \left( \frac{r - r_h}{r_h - r_c} \right)^{l'} + B \left( \frac{r_h - r_c}{r - r_h} \right)^{l'+1},
\end{equation}
with the coefficient ratio
\begin{equation}
\frac{B}{A} = \frac{\Gamma(-2 l' - 1) \Gamma(l' + 1) \Gamma(l' + 1 - 2i Q)}
{\Gamma(-l' - 2i Q) \Gamma(-l') \Gamma(2 l' + 1)}.
\end{equation}
Similarly, the far-region solution~\eqref{Sols:far-region_small-r} reads
\begin{equation}
R(r) \sim A' r^{l'} + B' r^{-l' - 1}, \quad
\frac{B'}{A'} = 2^{-2 l' - 1} \kappa^{-2 l' - 1} \frac{\Gamma(-l' - \nu)}{\Gamma(l' - \nu + 1)}.
\end{equation}
Equating the asymptotic expansions yields the reflection coefficient
\begin{equation}
|\mathcal{R}|^2 \propto \left| \frac{\Gamma(-l' - \nu)}{\Gamma(l' + 1 - \nu)} \right|^2.
\end{equation}
By using the identity $\frac{\Gamma(-2l'-1)}{\Gamma(-l')}=\frac{(-1)^{l'+1}l!}{2(2l'+1)!}$ (see details of proof in \cite{Bao2022}), one yields the corresponding low-frequency amplification factor to leading order when $\epsilon \ll 1$ and $l'\approx l$:
\begin{equation}
\label{Eq:Zlm-EMDA-PDFM}
Z_{lm}(\omega) = -4 Q (i \kappa)^{2 l + 1}
\frac{(l!)^4}{[(2 l)!]^2 [(2 l + 1)!]^2}
\prod_{j=1}^l \left( 1 + \frac{4 Q^2}{j^2} \right),
\end{equation}
where \( Q \) depends on \( r_2 \) and \( \lambda \) through the horizon locations.

\begin{figure}[h]
\centering
\includegraphics[width=\textwidth]{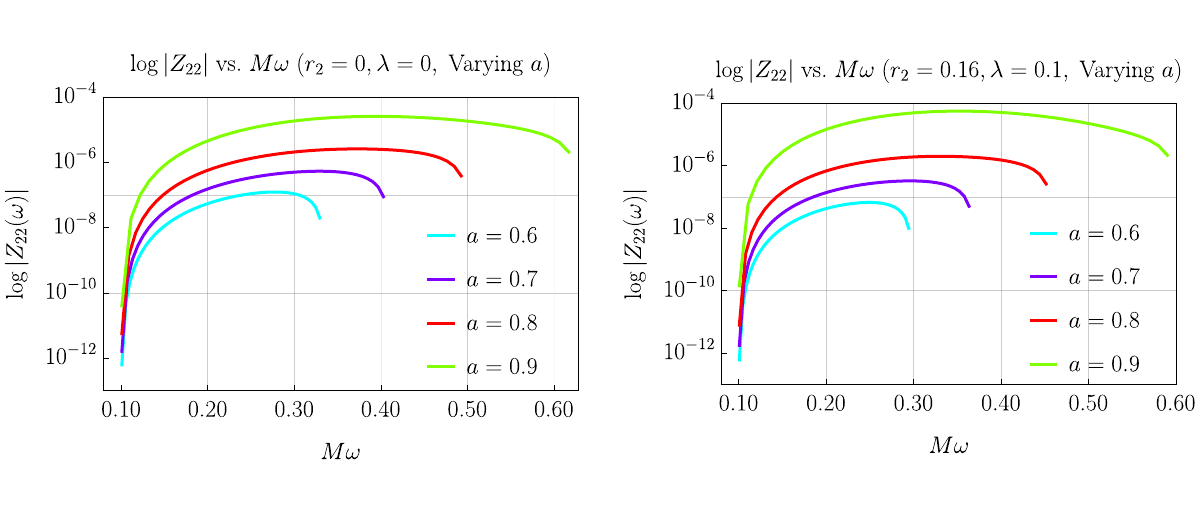}
\caption{Low-frequency amplification factor \( |Z_{22}| \) computed from Eq.~\eqref{Eq:Zlm-EMDA-PDFM} as a function of frequency \( \omega \) for a Kerr black hole (\( r_2 = 0 \), \( \lambda = 0 \)) (left) and a black hole with dilaton charge \( r_2 = 0.16 \) and PFDM parameter \( \lambda = 0.1 \) (right), with spins \( a = \{0.6, 0.7, 0.8, 0.9\} \). Parameters: \( M = 1 \), \( \mu = 0.1 \), \( l = 2 \), \( m = 2 \). Increasing spin enhances the superradiant amplification and broadens the corresponding frequency window.}
\label{fig:ZlmSpin}
\end{figure}

\begin{figure}[h]
\centering
\includegraphics[width=\textwidth]{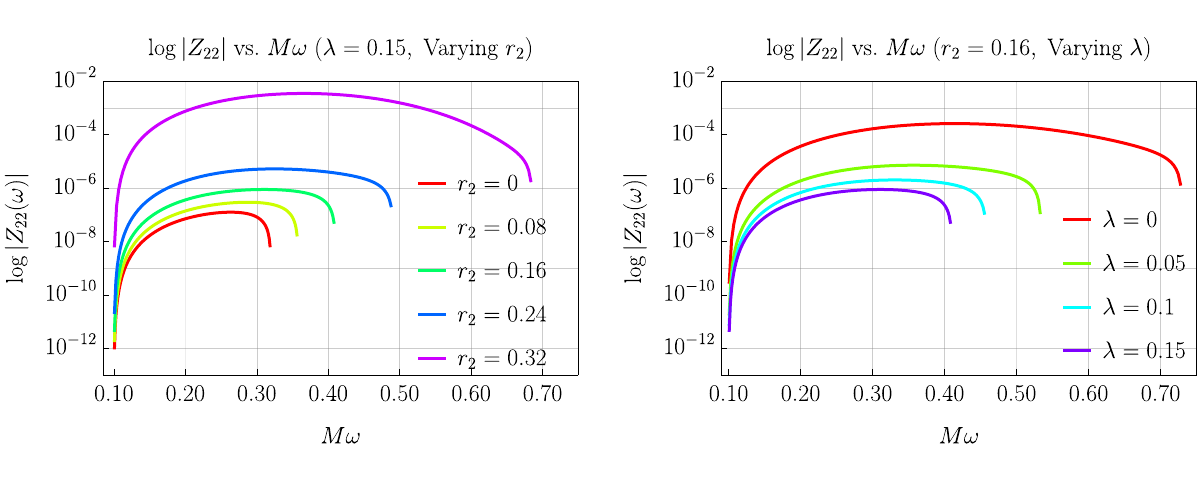}
\caption{Same as Fig.~\ref{fig:ZlmSpin}, but showing the dependence on the dilaton parameter \( r_2 \) and PFDM parameter \( \lambda \). Left: \( \lambda = 0.15 \), \( r_2 = \{0.0, 0.08, 0.16, 0.24, 0.32\} \). Right: \( r_2 = 0.16 \), \( \lambda = \{0.0, 0.05, 0.10, 0.15\} \). Parameters: \( M = 1 \), \( \mu = 0.1 \), \( l = 2 \), \( m = 2 \), \( a = 0.8 \).  Increasing the dilaton parameter \( r_2 \) enhances the superradiant amplification, whereas increasing the PFDM parameter \( \lambda \) suppresses it.}
\label{fig:Zlm}
\end{figure}

Figures~\ref{fig:ZlmSpin} and~\ref{fig:Zlm} present numerical evaluations of \( Z_{lm} \) obtained from the above analytic formula, illustrating the effects of spin, dilaton charge, and PFDM density. In the \(\lambda =0\) limit, our low-frequency results reproduce the qualitative behavior of Senjaya and Ponglertsakul~\cite{Senjaya2025}, who studied the pure EMDA case: larger spin \( a \) and dilaton parameter \( r_2 \) increase amplification and enlarge superradiant frequency window. For \(\lambda > 0\), we find a suppression of \( Z_{lm} \)--a stabilizing dark matter effect absent in pure EMDA. These results confirm that the amplification factor \( Z_{lm} \) is strongly sensitive to the black hole's spin and dilaton charge, while PFDM acts as a stabilizing agent by suppressing amplification and narrowing the superradiant frequency window.

\subsection{Energy Extraction from the Black Hole}
\label{subSec:Energy_Extraction}
In this subsection, we investigate the energy extraction from a Kerr black hole in the EMDA theory surrounded by PFDM, leveraging the superradiant amplification effects. Superradiance enables energy extraction from rotating black holes, offering insights into dark matter's influence on black hole dynamics. We compute the net extracted energy and analyze its dependence on the spin \( a \), PFDM parameter \( \lambda \), and dilaton parameter \( r_2 \).

For a monochromatic massless scalar field ($\mu=0$), the outgoing energy flux at spatial infinity is
\begin{equation}
\label{eq:outgoing_energy_flux}
\dot{E}_{\rm out}=\frac{\omega k_{\infty}}{2}|\mathcal{R}|^2,
\end{equation}
which follows directly from the scalar-field energy-momentum tensor (see Ref.~\cite{Liu2023superradiance} for the detailed derivation).

For non-monochromatic incident waves with a thermal spectrum at temperature $T\gg T_H$ (where $T_H$ is the Hawking temperature), spontaneous Hawking emission is negligible. The spectral weight is given by the normalized thermal (blackbody) mode density~\cite{Wondrak2018,Supakorn2026}
\begin{equation}
n(\omega)=\frac{\omega^2}{\mathcal{N}\bigl(\exp[\omega/k_B T]-1\bigr)},
\end{equation}
where the normalization constant $\mathcal{N}$ is fixed by the condition $\int_0^\infty n(\omega)\,d\omega=1$ together with the Riemann zeta integral~\cite{Abramowitz1970} 
\begin{equation}
\int_0^\infty\frac{x^{s-1}}{e^x-1}\,dx=\Gamma(s)\zeta(s)\qquad(\Re s>1).
\end{equation}
Evaluating the integral yields
\begin{equation}
\mathcal{N}=\Gamma(3)\zeta(3)(k_B T)^3\approx2.404(k_B T)^3.
\end{equation}

The total extracted energy rate is obtained by weighting the monochromatic flux with this spectrum:
\begin{equation}
\dot{E}_{\rm tot}=\int_0^\infty\dot{E}_{\rm out}\,n(\omega)\,d\omega.
\end{equation}
Subtracting the ingoing flux for each mode then gives the net extracted energy rate
\begin{equation}
\label{Eq:net_extracted_energy_rate}
\dot{E}_{\rm net}=\int_0^\infty\frac{\omega k_{\infty}}{2}\bigl(|\mathcal{R}|^2-|\mathcal{I}|^2\bigr)n(\omega)\,d\omega.
\end{equation}
This quantity is positive in the superradiant regime $0<\omega<m\Omega_h$ and negative for $\omega>m\Omega_h$. Here the horizon angular velocity
\begin{equation}
\Omega_h=\frac{a}{r_h(r_h+2r_2)+a^2}
\end{equation}
sets the critical frequency (scaled by $m$)~\cite{Wondrak2018}; it encodes the full dependence on the spin parameter $a$, the dilaton parameter $r_2$, and the PFDM strength $\lambda$ through the event-horizon radius $r_h$ (the outer root of $\Delta_D(r)=0$).

\begin{figure}[h]
    \centering
    \includegraphics[width=0.8\textwidth]{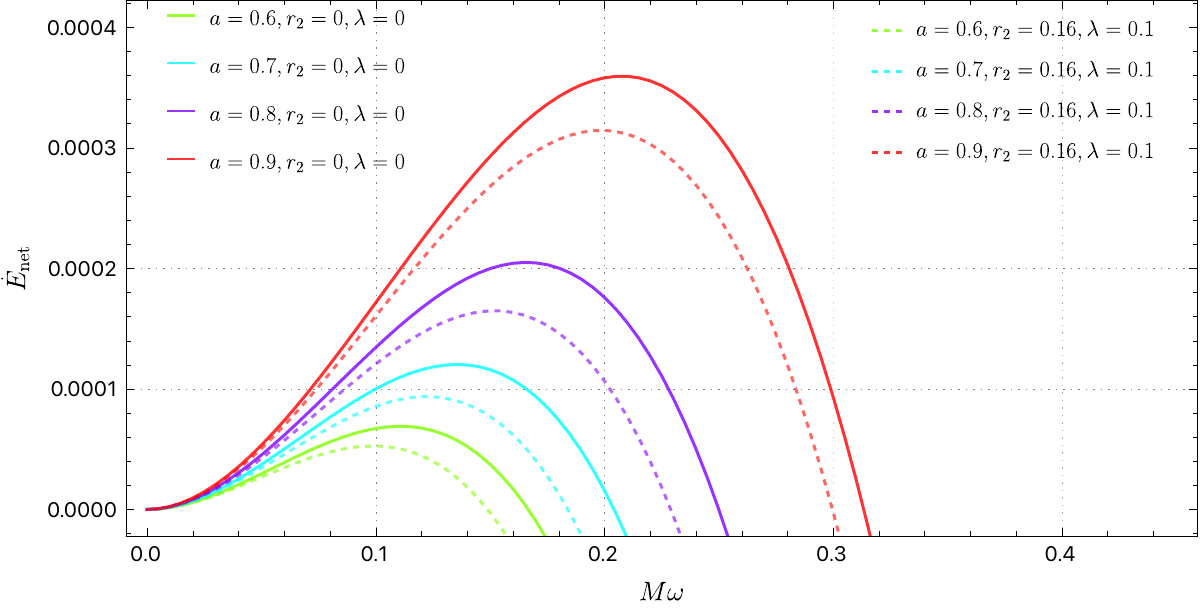}
    \caption{Net extracted energy varying \( a \) for fixed \( r_2 \) and \( \lambda \), showing increased extraction with higher \( a \).}
  \label{NetExtractedEnergy_varying_a}
\end{figure}

\begin{figure}[h]
    \centering
    \includegraphics[width=0.8\textwidth]{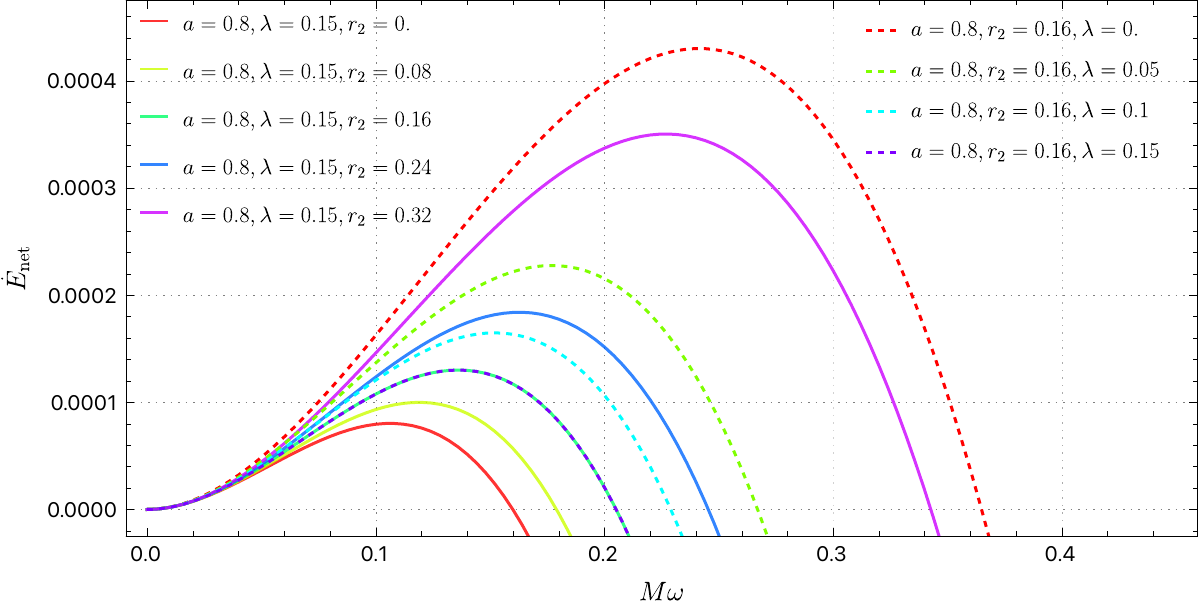}
    \caption{Net extracted energy varying \( r_2 \) and \( \lambda \) for fixed \( a \), demonstrating enhancement with \( r_2 \) and suppression with \( \lambda \).}
    \label{NetExtractedEnergy_varying_r2_lambda}
\end{figure}

Although Eq.~\eqref{Eq:net_extracted_energy_rate} is formally defined over $0\leq\omega<\infty$, the numerical evaluation shown in Figs.~\ref{NetExtractedEnergy_varying_a} and \ref{NetExtractedEnergy_varying_r2_lambda} is performed only in the superradiant sector. Because the integrand is positive solely for $0<\omega<m\Omega_h$ and changes sign above this threshold, we integrate from $\omega=0$ to a value slightly above $m\Omega_h$ (typically $m\Omega_h+0.01$) in order to confirm the sign transition. All results are for the dominant mode $l=m=1$ (strongest superradiant amplification~\cite{Brito2020,Press1972}). We work in natural units with $k_B=1$ and adopt the cosmic-microwave-background temperature $T=2.7\,\text{K}$. 

Figure~\ref{NetExtractedEnergy_varying_a} shows that energy extraction increases with the spin parameter $a$ (at fixed $r_2$ and $\lambda$) because the ergoregion enlarges. Figure~\ref{NetExtractedEnergy_varying_r2_lambda} demonstrates that larger $r_2$ enhances extraction while larger $\lambda$ suppresses it, fully consistent with the behaviour of the amplification factor $Z_{lm}$.

The ergosphere, between infinite redshift surface \( r_{rs} \) and \( r_h \), is approximated by solving \( g_{tt} = 0 \) up to first-order correction~\cite{Li2023}:
\begin{equation}
r_{rs} \equiv r_{rs}^{(1)} \approx (M - r_2) + \sqrt{(M - r_2)^2 - a^2 \cos^2\theta + (\lambda/2) \ln\left(r_{rs}^{(0)}/\lambda\right)},
\end{equation}
with \( r_h \) from \( \Delta_D(r_h) = 0 \). The size \( r_{\text{ergo}} = r_{rs} - r_h \) increases with \( a \), decreases with \( \lambda \), and modulates with \( r_2 \), consistent with simulations (Tables~\ref{Ergosphere_varying_a}–\ref{Ergosphere_varying_r2_lambda}). These results highlight the interplay between rotation, dark matter density, and dilaton contributions in EMDA-PFDM black holes, laying groundwork for observational constraints on dark matter profiles.

\begin{table}[h]
\centering
\caption{Ergosphere and net extracted energy for fixed \(r_2\) and \(\lambda\), and varying \(a\).}
\begin{tabular}{cccccc} \hline
\multicolumn{6}{c}{Case I: \( r_2 = 0 \), \( \lambda = 0 \), varying \(a\)} \\
\hline
$a$ & $r_{\text{rs}}$ & $r_h$ & $r_{\text{ergo}}$ & $r_{\text{rs}}/r_h$ & $E_{\text{net}}\times 10^{-4}$ \\
\hline
0.60 & 2.00000 & 1.80000 & 0.200004 & 1.11111 & $0.064820$ \\
0.70 & 2.00000 & 1.71414 & 0.285861 & 1.16677 & $0.138469$ \\
0.80 & 2.00000 & 1.60000 & 0.400004 & 1.25000 & $0.288984 $ \\
0.90 & 2.00000 & 1.43589 & 0.564114 & 1.39287 & $0.635917 $ \\
\hline
\multicolumn{6}{c}{Case II: \( r_2 = 0.16 \), \( \lambda = 0.1 \), varying \(a\)} \\
\hline
$a$ & $r_{\text{rs}}$ & $r_h$ & $r_{\text{ergo}}$ & $r_{\text{rs}}/r_h$ & $E_{\text{net}}\times 10^{-4}$\\
\hline
0.60 & 1.97847 & 1.76271 & 0.215759 & 1.12240 & $0.044214 $ \\
0.70 & 1.97847 & 1.66755 & 0.310923 & 1.18646 & $0.097156 $ \\
0.80 & 1.97847 & 1.53677 & 0.441704 & 1.28742 & $0.212879 $ \\
0.90 & 1.97847 & 1.32947 & 0.649000 & 1.48816 & $0.532040 $ \\
\hline
\end{tabular}
\label{Ergosphere_varying_a}
\end{table}
\begin{table}[h]
\centering
\caption{Ergosphere and net extracted energy for a fixed \(a\), and varying \(r_2\) and \(\lambda\), respectively.}
\begin{tabular}{cccccc} \hline
\multicolumn{6}{c}{Case I: \( a = 0.8 \), \( \lambda = 0.15 \), varying \(r_2\)} \\
\hline
$r_2$ & $r_{\text{rs}}$ & $r_h$ & $r_{\text{ergo}}$ & $r_{\text{rs}}/r_h$ &  $E_{\text{net}}\times 10^{-4}$\ \\
\hline
0.00 & 2.41694 & 2.08862 & 0.328322 & 1.15720 & $0.072391$ \\
0.08 & 2.24593 & 1.87842 & 0.367505 & 1.19565 & $0.100854 $ \\
0.16 & 2.07399 & 1.65268 & 0.421312 & 1.25493 & $0.150469 $ \\
0.24 & 1.90092 & 1.39628 & 0.504644 & 1.36142 & $0.254755 $ \\
0.32 & 1.72648 & 1.02037 & 0.706114 & 1.69202 & $0.678139 $ \\
\hline
\multicolumn{6}{c}{Case II: \( a = 0.8 \), \(r_2 = 0.16 \), varying \( \lambda\)} \\
\hline
$\lambda$ & $r_{\text{rs}}$ & $r_h$ & $r_{\text{ergo}}$ & $r_{\text{rs}}/r_h$ &  $E_{\text{net}}\times 10^{-4}$\ \\
\hline
0.00 & 1.68000 & 1.09613 & 0.583875 & 1.53267 & $0.886723 $ \\
0.05 & 1.86084 & 1.38338 & 0.477464 & 1.34514 & $0.342947 $ \\
0.10 & 1.97847 & 1.53677 & 0.441704 & 1.28742 & $0.212879 $ \\
0.15 & 2.07399 & 1.65268 & 0.421312 & 1.25492 & $0.150469 $ \\
\hline
\end{tabular}
\label{Ergosphere_varying_r2_lambda}
\end{table}

\subsection{Superradiant Instability}
\label{Sec:Superradiant_Instability}
For massive scalar fields around rotating black holes, the scalar mass acts as a natural mirror, enabling the superradiant instability when the mode frequency satisfies both \(\omega < m \Omega_h\) and \(\omega < \mu\) \cite{Myung2022}. In this regime, low-frequency modes (\(\omega < \mu\)) are trapped and cannot propagate to infinity. Combined with superradiance, this confinement can lead to exponentially growing modes. The effective potential typically exhibits a barrier (local maximum) and a trapping well (local minimum) outside the ergoregion. Waves reflected by the barrier are amplified in the well, leading to an instability. The boundary conditions for such unstable modes are
\begin{equation}
\label{Sols:Schrödinger-like_instability}
\Psi(r_*) \sim 
\begin{cases} 
e^{-i (\omega - m \Omega_h) r_*}, & r_* \to -\infty \, (r \to r_h), \\ 
e^{-\kappa r_*}, & r_* \to \infty \, (r \to \infty),
\end{cases}
\end{equation}
ensuring purely outgoing behavior at the outer horizon and exponential decay at infinity. 

To assess the impact of PFDM, we consider the radial equation~\eqref{Eq:radial_part}. Defining \( R(r) = \Psi(r) / \sqrt{\Delta_D(r)} \) recasts it into a Schrödinger-like form in the radial coordinate:
\begin{equation}
\Psi''(r) + V(r) \Psi(r) = 0,
\end{equation}
with
\begin{equation}
\label{Def:V_eff}
\begin{aligned}
V(r) = & \frac{\left( \omega \left[ r (r + 2 r_2) + a^2 \right] - a m \right)^2}{\Delta_D(r)^2} - \frac{\left( K_{lm} + r(r + 2 r_2) \mu^2 \right) \Delta_D(r)}{\Delta_D(r)^2} \\
& + \frac{\Delta_D'(r)^2 - 2 \Delta_D(r) \Delta_D''(r)}{4 \Delta_D(r)^2},
\end{aligned}
\end{equation}
where \( K_{lm} = \Lambda_{lm} - 2 m a \omega + a^2 \omega^2 \).

For large \(r\), the effective potential behaves as
\begin{equation}
\label{Def:V_eff}
V_{\text{eff}}(r) = \omega^2 - V(r) \approx \mu^2 + \frac{(\mu^2 - 2 \omega^2) \left[ 2 M + \lambda \ln \left( \frac{r}{\lambda} \right) \right]}{r} + \mathcal{O}\!\left( \frac{1}{r^2} \right),
\end{equation}
with its radial derivative
\begin{equation}
\label{Def:Derivative_V_eff}
V_{\text{eff}}'(r) \approx -\frac{(\mu^2 - 2 \omega^2) \left[ 2 M - \lambda + \lambda \ln \left( \frac{r}{\lambda} \right) \right]}{r^2}.
\end{equation}

The asymptotic form of the effective potential~\eqref{Def:V_eff} and its derivative~\eqref{Def:Derivative_V_eff} provide a useful heuristic for the formation of trapping wells. When \(\omega > \mu/\sqrt{2}\), the \(1/r\) coefficient in \(V_{\rm eff}(r)\) becomes negative, producing a positive asymptotic slope \(V'_{\rm eff}(r) > 0\) at large \(r\). This suggests the potential may develop a local minimum (trapping well) at finite radius, which is a necessary (but not sufficient) condition for quasi-bound states. Full superradiant instability additionally requires the superradiance condition \(\omega < m\Omega_H\) and the bound-state condition \(\omega < \mu\).

The PFDM parameter \(\lambda\) enters the effective potential through its logarithmic contribution and, according to our numerical results, deepens the trapping well. Nevertheless, the superradiant amplification factor \(Z_{lm}\) and the extracted rotational energy both decrease with increasing \(\lambda\). This indicates that, in the EMDA black-hole background with PFDM, the suppression of horizon energy extraction dominates over the enhanced trapping, producing a net stabilizing tendency.

No sharp critical value of \(\lambda\) or \(r_2\) is found within the physically allowed parameter space. The trapping well persists for all \(\lambda/M \lesssim 0.15\) and for all \(r_2\) values that still yield regular event and Cauchy horizons (i.e., real roots of \(\Delta_D(r)=0\)). The upper bound on \(r_2\) is therefore set by the black-hole condition itself rather than by the disappearance of the trapping well.

\begin{figure}[h]
    \centering
    \includegraphics[width=0.9\textwidth]{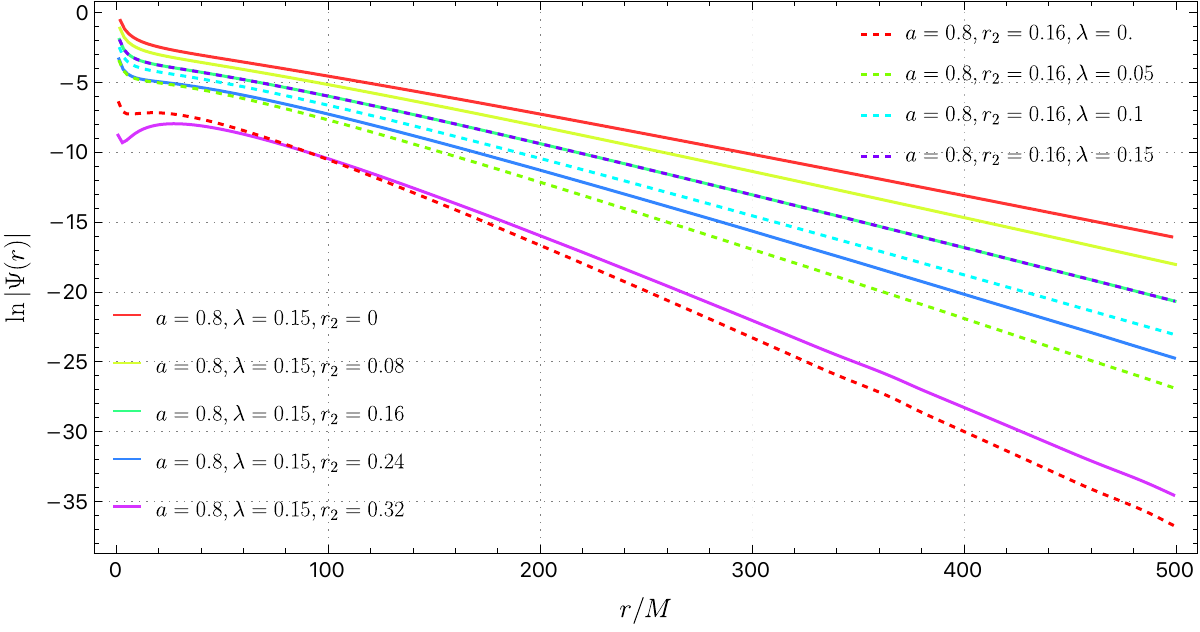}
    \caption{%
    Approximate quasi-bound-state scalar profiles at large radius obtained from the far-region radial equation~\eqref{Eq:Far_radial_approx} near the superradiant threshold (\(\operatorname{Re}(\omega)\approx m\Omega_H\), \(\operatorname{Im}(\omega)=0\)). 
    The linear behaviour of \(\ln|\Psi(r)|\) confirms the expected exponential decay. 
    Increasing the dilaton parameter \(r_2\) steepens the decay (stronger localization), while increasing the PFDM parameter \(\lambda\) flattens it (weaker localization).}
    \label{Approximate_solutions_varying_r2_lambda_Combined}
\end{figure}

To quantify the interplay between trapping and superradiant amplification, we examine the asymptotic localization of scalar-cloud-type solutions near the superradiant threshold. We numerically solve the far-region radial equation~\eqref{Eq:Far_radial_approx} (incorporating the geometric coefficients~\eqref{eqs:geometric_coefficients}) subject to exponentially decaying boundary conditions at large \(r\), choosing \(\operatorname{Re}(\omega)\approx m\Omega_H\) and \(\operatorname{Im}(\omega)=0\)~\cite{Senjaya2025}. The resulting profiles (Fig.~\ref{Approximate_solutions_varying_r2_lambda_Combined}) exhibit the expected exponential decay \(\ln|\Psi(r)| = -\kappa r + c\) for \(r \gtrsim 350M\).

We define the effective localization length
\[
L_{\rm loc,num} = \frac{1}{\kappa_{\rm num}},
\]
where \(\kappa_{\rm num}\) is extracted from a linear fit in the large-radius region. As shown in Table~\ref{Tab:Effective_length_varying_r2_lambda}, the numerical values agree with the theoretical decay constant \(\kappa_{\rm th}\) within \(6\)--\(8\,\%\). The trends are clear: increasing \(r_2\) (at fixed \(\lambda\)) raises \(\kappa_{\rm num}\) and shortens \(L_{\rm loc,num}\) (more compact cloud), while increasing \(\lambda\) (at fixed \(r_2\)) lowers \(\kappa_{\rm num}\) and lengthens \(L_{\rm loc,num}\) (more extended cloud). These localization properties are consistent with the observed suppression of energy extraction by PFDM and support the picture of a competition between horizon amplification and spatial confinement of the scalar field.
\begin{table}[h]
\centering
\caption{%
Numerical and theoretical exponential decay constants \(\kappa\) and effective localization lengths \(L_{\rm loc}=1/\kappa\) for asymptotic scalar-cloud-type solutions near the superradiant threshold. 
Numerical values \(L_{\rm loc,num}\) are extracted from linear fits \(\ln|\Psi(r)|=-\kappa r + c\) in the region \(r\geq 350M\); theoretical values follow from \(\kappa_{\rm th}\).}
\label{Tab:Effective_length_varying_r2_lambda}
\begin{tabular}{cccccc} \hline
\multicolumn{6}{c}{Case I: \( a = 0.8 \), \( \lambda = 0.15 \), varying \(r_2\)} \\
\hline
$r_2$ & $ \kappa_{\rm num}$ & $\kappa_{\rm th} $ & $L_{\rm loc,num}$ & $L_{\rm loc,th}$ & $\Delta L_{\rm loc}\,(\%)$ \\
\hline
0 &  0.02996  &  0.03199  &  33.376  &  31.265  &  6.75   \\
0.08 &  0.03350  &  0.03580  &  29.854  &  27.931  &  6.88  \\
0.16 &  0.03834  &  0.04102  &  26.082  &  24.376  &  7.00  \\
0.24 &  0.04579  &  0.04908  &  21.839  &  20.374  &  7.19  \\
0.32 &  0.06335  &  0.06855  &  15.786  &  14.589  &  8.21 \\
\hline
\multicolumn{6}{c}{Case II: \( a = 0.8 \), \(r_2 = 0.16 \), varying \( \lambda\)} \\
\hline
$\lambda$ & $ \kappa_{\rm num}$ & $\kappa_{\rm th} $ & $L_{\rm loc,num}$ & $L_{\rm loc,th}$ & $\Delta L_{\rm loc}\,(\%)$ \\
\hline
0.00 &  0.06815  &  0.07298  &  14.673  &  13.702  &  7.09 \\
0.05 &  0.05004  &  0.05340  &  19.982  &  18.728  &  6.70 \\
0.10 &  0.04287  &  0.04580  &  23.326  &  21.834  &  6.83  \\
0.15 &  0.03834  &  0.04102  &  26.082  &  24.376  &  7.00   \\
\hline
\end{tabular}
\end{table}

\section{Quasi-bound States of a Massive Scalar Field in Kerr--EMDA--PFDM Spacetime}
\label{Sec:Quasi-bound_States}

We extend the analysis of quasi-bound states for a massive scalar field in the Kerr background, as originally presented by Detweiler~\cite{detweiler1980black}, to the more general case of a Kerr black hole in the Einstein–Maxwell–dilaton–axion (EMDA) theory surrounded by perfect fluid dark matter (PFDM). The dilaton parameter \(r_2\) and PFDM parameter \(\lambda\) modify the background metric, altering the effective potential in the scalar field radial equation and thereby shifting the quasi-bound state spectrum. These effects are particularly relevant for ultralight bosons, for which a hydrogen-like approximation applies. 

A distinction is made between \emph{pure bound states}, characterized by discrete real frequencies (neglecting black hole absorption), and \emph{quasi-bound states}, which possess complex frequencies due to coupling to the horizon and may exhibit superradiant instability. Quasi-normal modes, corresponding to purely outgoing boundary conditions at infinity and describing black hole ringdown, will be addressed in the subsequent section. For reference, the exact quasi-bound states and scalar clouds of Kerr--EMDA black holes (\(\lambda=0\)) have been studied in detail by Senjaya and Ponglertsakul~\cite{Senjaya2025} using confluent Heun function techniques; our results reduce to theirs in the appropriate limit.

\subsection{Review of the Standard Kerr Case}
For a massive scalar field in the Kerr spacetime, the far-region radial equation simplifies for \( r \gg M \) to a Schrödinger-like form~\cite{detweiler1980black}:
\begin{equation}
\label{Eq:Far_radial_approx2}
\Psi''(r) + \left[ -\kappa^2 + \frac{2 \kappa \nu}{r} - \frac{l'(l'+1)}{r^2} \right] \Psi(r) = 0,
\end{equation}
where \(\Psi(r) = r R(r)\), \(\kappa^2 = \mu^2 - \omega^2\), \(\nu = (M\mu^2 - 2 M\omega^2) / \kappa\), and \(l'\) is a shifted angular momentum index (\(l' \approx l\) for \(a\mu \ll 1\)). The far-region potential is
\begin{equation}
\label{Eq:Far_Potential}
V_{\text{Far}}(r) = -\kappa^2 + \frac{2 \kappa \nu}{r} - \frac{l'(l'+1)}{r^2}.
\end{equation}
Introducing \( z = 2\kappa r \) and the ansatz
\begin{equation}
\label{Eq:sol_radial_approx2}
\Psi(z) = e^{-z/2} z^{l'+1} \psi(z),
\end{equation}
transforms Eq.~\eqref{Eq:Far_radial_approx2} into the Whittaker form of the confluent hypergeometric equation:
\begin{equation}
z \frac{d^2 \psi}{dz^2} + \left[ 2(l'+1) - z \right] \frac{d\psi}{dz} + \left( \nu - (l'+1) \right) \psi = 0.
\end{equation}
The general solution is
\begin{equation}
\psi(z) = C_1\, M(1 - \nu + l', 2 + 2l', z) + C_2\, U(1 - \nu + l', 2 + 2l', z),
\end{equation}
where \(M\) and \(U\) are confluent hypergeometric functions of the first and second kind, respectively.

\subsubsection{Bound States}
Pure bound states require \(\omega < \mu\) (real \(\omega\)) so that the solution decays exponentially at spatial infinity. The large-\(z\) behavior shows that the \(M\) function generally grows exponentially unless its series terminates. Imposing the termination condition
\begin{equation}
1 - \nu + l' = -n_r, \quad n_r \in \mathbb{Z}_{\ge 0},
\end{equation}
ensures polynomial truncation, yielding normalizable solutions. For slow rotation \(l'\approx l\) this gives a quantization condition for principal quantum number
\begin{equation}
n \equiv \nu  \simeq n_r + l + 1,
\end{equation}
leading to the hydrogen-like spectrum
\begin{equation}
\omega \approx \mu - \frac{M^2 \mu^3}{2 n^2},
\end{equation}
valid for \(\mu M \ll 1\) and neglecting \(\mathcal{O}((\mu M)^4)\) and rotational corrections.

\subsubsection{Quasi-bound States}
Quasi-bound states have \(\mathrm{Re}(\omega) < \mu\) but \(\mathrm{Im}(\omega) \ne 0\), with \(\mathrm{Im}(\omega) > 0\) indicating instability. At infinity, the decaying behavior \(\Psi(r) \sim e^{-\kappa r}\) is obtained by setting \(C_1 = 0\), since the \(U\) function yields the correct asymptotics without polynomial termination:
\begin{equation}
\psi(z) = U(1 - \nu + l', 2 + 2l', z).
\end{equation}
Matching this far-region solution to the near-horizon ingoing wave condition determines the complex spectrum. In the small-\(\mu M\) limit, Detweiler~\cite{detweiler1980black} found that the superradiant instability growth time scales as \(\tau \sim (M/\mu)^{24} M / \mu^9\).

\subsection{Modifications from \(r_2\) and \(\lambda\)}
In the Kerr--EMDA--PFDM spacetime, the far-region potential acquires additional terms:
\begin{equation}
V_{\text{Far}}(r) \approx -\kappa^2 + \frac{2 \kappa \nu_{\rm eff}}{r} + \frac{\lambda \mu^2 \ln(r / \lambda)}{r} - \frac{l'(l'+1)}{r^2},
\end{equation}
where
\begin{equation}
\nu_{\rm eff} = \frac{(M - r_2)(\mu^2 - 2\omega^2)}{\kappa}
\end{equation}
reflects the effective mass shift \(M_{\rm eff} = M - r_2\) due to the dilaton. The logarithmic PFDM term acts as a perturbative correction.

\subsubsection{Bound States}
With \(\lambda = 0\), the hydrogen-like spectrum shifts to
\begin{equation}
\omega \approx \mu - \frac{(M - r_2)^2 \mu^3}{2 n^2}.
\end{equation}
This agrees with the small-\(\mu M\) limit of the EMDA result in Ref.~\cite{Senjaya2025}. For small but nonzero \(\lambda\), first-order perturbation theory gives
\begin{equation}
\delta_1 = \lambda \mu^2 \left\langle \frac{\ln(r / \lambda)}{r} \right\rangle,
\end{equation}
where the expectation value is evaluated over the unperturbed hydrogenic state. For the ground state (\(n=1\), \(l=0\)):
\begin{equation}
\left\langle \frac{\ln(r / \lambda)}{r} \right\rangle = (M - r_2)\mu^2 \left[ 1 - \gamma - \ln\left( 2(M - r_2)\mu^2 \right) - \ln \lambda \right],
\end{equation}
yielding
\begin{equation}
\omega \approx \mu - \frac{(M - r_2)^2 \mu^3}{2 n^2} - \lambda \mu^2 \left\langle \frac{\ln(r / \lambda)}{r} \right\rangle.
\end{equation}
Positive \(\lambda\) reduces the binding energy, a feature absent in the pure EMDA case.

\subsubsection{Quasi-bound States}
For quasi-bound states, the far-region decaying \(U\) solution is retained, with \(\nu \to \nu_{\rm eff}\). The dilaton increases the effective coupling, potentially enhancing the instability for fixed \(a\). In the \(\lambda \to 0\) limit, the far-region solution and matching conditions reproduce the EMDA results of Ref.~\cite{Senjaya2025}. For small but finite \(\lambda\), the PFDM term perturbs the growth rate, generally suppressing the instability by raising the minimum of \(V_{\text{eff}}\), weakening the confinement, and reducing the depth of the trapping well. This trend is consistent with the suppression observed in the superradiant amplification factor (Sec.~\ref{Matching_Leading-Order_Terms}).

\section{Quasinormal Modes of Kerr--EMDA--PFDM Black Hole}
\label{sec:QNM-EMDA-PFDM}

Quasinormal modes (QNMs) are the characteristic damped oscillations of black holes, described by complex frequencies \(\omega=\Re\omega+i\,\Im\omega\) whose real and imaginary parts set the oscillation rate and damping time, respectively. They are defined by boundary conditions of purely ingoing waves at the event horizon and purely outgoing radiation at spatial infinity~\cite{Regge1957,Vishveshwara1970a,Vishveshwara1970b,Chandrasekhar1983}. For a massive scalar field, the far-field solution decays exponentially and the problem connects continuously to quasi-bound states. QNM spectrum depend only on black-hole parameters and the underlying gravitational theory, thereby encoding direct information about the spacetime geometry.

In what follows we compute QNMs of a massive scalar field in a Kerr spacetime modified by Einstein–Maxwell–dilaton–axion (EMDA) theory and surrounded by perfect fluid dark matter (PFDM). The calculation is implemented in Mathematica using the asymptotic iteration method (AIM)~\cite{Ciftci2003,Cho2009,Cho2010,Ponglertsakul2019,Daghigh2020}.

\subsection{Implementation of the Asymptotic Iteration Method}

The Kerr--EMDA--PFDM metric in Boyer--Lindquist coordinates is summarized in Sec.~\ref{sec:EMDA_Theory_with_PFDM}, with the radial function \(\Delta_D(r)\) defined in Eq.~\eqref{Def:rho2_DeltaD}. Horizons are located at the roots of \(\Delta_D(r)=0\): the outer horizon \(r_h\) and inner (Cauchy) horizon \(r_c<r_h\). The horizon angular velocity and surface gravity are
\begin{equation}
\Omega_h = \frac{a}{r_h (r_h + 2 r_2) + a^2}, \qquad
\kappa_h = \frac{1}{2\!\left[ r_h (r_h + 2 r_2) + a^2 \right]}\,
\left.\frac{d\Delta_D}{dr}\right|_{r=r_h},
\end{equation}
reducing to the Kerr expressions when \(r_2=\lambda=0\).

A scalar field \(\Phi\) of mass \(\mu\) obeys the Klein--Gordon equation \eqref{Eq:Klein-Gordon}. With the separation ansatz
\begin{equation}
\Phi(t,r,\theta,\phi)=e^{-i\omega t} e^{i m \phi}\, S(\theta)\, R(r),
\end{equation}
one obtains the angular and radial equations \eqref{Eq:angular_part} and \eqref{Eq:radial_part}. For \(a=0\), the separation constant reduces to \(\Lambda_{lm}=l(l+1)\).
To render the radial equation into AIM form, we introduce the monotonic compactifying map
\begin{equation}
\xi(r)=\frac{r-r_h}{\,r-r_c\,}, \qquad r(\xi)=\frac{r_h-\xi\, r_c}{1-\xi}, \qquad 
\frac{d\xi}{dr}=\frac{r_h-r_c}{(r-r_c)^2},
\end{equation}
which sends \(r=r_h\to\xi=0\) and \(r\to\infty\to\xi\to 1\).
We factor out the near-horizon and far-field behaviors with
\begin{equation}
\label{eq:radial_ansatz}
R(r)=e^{\kappa r}\, (r-r_c)^{i\sigma_+ + \kappa -1}\, (r-r_h)^{-i\sigma_+}\, \Psi(\xi),
\end{equation}
where
\begin{equation}
\sigma_+=\frac{\omega-m\Omega_h}{2\kappa_h}
=\frac{\big[r_h(r_h+2r_2)+a^2\big](\omega-m\Omega_h)}{\,r_h-r_c\,},
\end{equation}
and \(\kappa=\sqrt{\mu^2-\omega^2}\) specifies the large-\(r\) behavior. 
For the QNM branch studied here we have \(\omega^2>\mu^2\), so \(\kappa=i k\) with \(k=\sqrt{\omega^2-\mu^2}>0\), and the factor \(\exp(\kappa r)=e^{+i k r}\) enforces outgoing radiation at infinity (up to \(1/r\) corrections). 
If one targets quasi-bound states (\(\omega^2<\mu^2\)), the decaying boundary condition is obtained by taking \(\kappa\to +|\kappa|\) and replacing \(e^{\kappa r}\) with \(e^{-|\kappa| r}\).

After the coordinate transformation \(r\mapsto \xi\), the radial equation \eqref{Eq:radial_part} becomes
\begin{equation}
\Delta_D(r(\xi)) \frac{d\xi}{dr}\,\frac{d}{d\xi}\!\left[ \frac{d\xi}{dr}\,\frac{d R\!\big(r(\xi)\big)}{d\xi} \right]
+\Delta_D'(r(\xi)) \frac{d\xi}{dr}\,\frac{d R\!\big(r(\xi)\big)}{d\xi}
+V\!\big(r(\xi)\big)\,R\!\big(r(\xi)\big)=0,
\end{equation}
where
\begin{equation}
V(r)=\frac{\big[\omega\big(r(r+2r_2)+a^2\big)-am\big]^2}{\Delta_D(r)}
-\Big[\Lambda_{lm}+\mu^2 r(r+2r_2)-2am\omega+a^2\omega^2\Big].
\end{equation}
With the ansatz \eqref{eq:radial_ansatz} and after collecting terms, the equation reduces to the standard AIM form for \(\Psi(\xi)\),
\begin{equation}
\frac{d^2\Psi}{d\xi^2}+\lambda_0(\xi)\frac{d\Psi}{d\xi}+s_0(\xi)\,\Psi=0,
\end{equation}
where, by construction,
\begin{equation}
\lambda_0(\xi)=-\frac{\text{coeff.\ of }\,\frac{d\Psi}{d\xi}}{\text{coeff.\ of }\,\frac{d^2\Psi}{d\xi^2}},
\qquad
s_0(\xi)=-\frac{\text{coeff.\ of }\,\Psi}{\text{coeff.\ of }\,\frac{d^2\Psi}{d\xi^2}}.
\end{equation}

\subsection{Numerical Procedure}

The coefficients \(\lambda_0(\xi)\) and \(s_0(\xi)\) are expanded in Taylor series about \(\bar{\xi}=0.3\):
\begin{equation}
\lambda_0(\xi) = \sum_{i=0}^{\infty} c_{0,i} (\xi - \bar{\xi})^i, \quad
s_0(\xi) = \sum_{i=0}^{\infty} d_{0,i} (\xi - \bar{\xi})^i.
\end{equation}
Higher-order coefficients are obtained iteratively:
\begin{align}
c_{n,i} &= (i+1) c_{n-1,i+1} + d_{n-1,i} + \sum_{k=0}^i c_{0,k} \, c_{n-1,i-k}, \\
d_{n,i} &= (i+1) d_{n-1,i+1} + \sum_{k=0}^i d_{0,k} \, c_{n-1,i-k}.
\end{align}
After \(N_{\mathrm{max}}\) iterations, the characteristic equation
\begin{equation}
\label{Eq:characteristic_AIM}
d_{N_{\mathrm{max}},0} c_{N_{\mathrm{max}}-1,0} - d_{N_{\mathrm{max}}-1,0} c_{N_{\mathrm{max}},0} = 0
\end{equation}
is solved numerically using \texttt{NSolve} with 50-digit precision. Solutions are selected to satisfy \(\mathrm{Im}(\omega) < 0\) (damped modes) and physically reasonable \(\mathrm{Re}(\omega)\).

\paragraph*{Angular eigenvalue.}
The angular function \(S(u)\) in  Eq.~\eqref{Equ:angular_part} satisfies the (oblate) spheroidal equation~\cite{Abramowitz1970}
\begin{equation}
\label{eq:AngularEq}
\frac{d}{du} \left[ (1 - u^2) \frac{d S}{du} \right]
+ \left[\Lambda_{lm}(c)+c^{2} u^2 - \frac{m^2}{1 - u^2} \right] S = 0,
\end{equation}
with complex spheroidal parameter \(c\).
For a massless scalar (\(\mu=0\)) we take \(c=a\,\omega\); for a massive scalar (\(\mu\neq 0\)) we use
\(c=a\sqrt{\omega^{2}-\mu^{2}}\).
The dilaton/PFDM parameters (\(r_{2},\lambda\)) affect only the radial sector and do not enter Eq.~\eqref{eq:AngularEq}. By mapping \(c \mapsto -i\gamma, u \mapsto z ,S  \mapsto y\), Eq~\eqref{eq:AngularEq} can be rearranged to exactly match with the Mathematica's build-in differential equation
\begin{equation}
\label{eq:build-in_angular}
(1-z^2)y^{\prime\prime}-2zy^{\prime}+\left(\ \mathrm{SpheroidalEigenvalue}[\,l,\,m,\, \gamma\,] +\gamma^2(1-z^2)-\frac{m^2}{1-z^2}\right)y=0.
\end{equation}

\paragraph*{Practical evaluation of the angular eigenvalue.}
We implemented an AIM solver for Eq.~\eqref{eq:AngularEq} and verified that, for complex oblate parameter \(c\), it reproduces the eigenvalues returned by \texttt{SpheroidalEigenvalue} to high accuracy across the parameter range considered. To keep the QNM pipeline simple and robust, we therefore evaluate the separation constant with Mathematica’s built-in routine via the oblate–prolate mapping,
\begin{equation}
\label{eq:LambdaFromMathematica}
\Lambda_{lm}(c)
= \mathrm{SpheroidalEigenvalue}[\,l,\,m,\, i\,c\,] \;-\; c^{2},
\qquad
c =
\begin{cases}
a\,\omega, & \mu=0,\\[2pt]
a\sqrt{\omega^{2}-\mu^{2}}, & \mu\neq 0,
\end{cases}
\end{equation}
where \(\omega\) is complex and the principal branch of the square root is used. The subtraction of \(c^{2}\) aligns Mathematica’s convention with Eq.~\eqref{eq:AngularEq}, ensuring that the radial potential appears in the combination \(\Lambda_{lm} - 2 a m \omega + a^{2}\omega^{2}\) as written in \(V(r)\).

\paragraph*{Two-pass update of the angular separation constant.}
To consistently account for the coupling between the angular and radial sectors, we adopt a two-pass procedure for \(\Lambda_{lm}\).
In the first pass we set \(\Lambda_{lm}\approx l(l+1)\) to obtain a provisional frequency \(\omega^{(0)}\) from the radial AIM.
We then form \(c^{(0)}=a\sqrt{\omega^{(0)2}-\mu^{2}}\) (reducing to \(a\omega^{(0)}\) when \(\mu=0\)) and evaluate \(\Lambda_{lm}(c^{(0)})\) using Eq.~\eqref{eq:LambdaFromMathematica}.
In the second pass we rerun the radial AIM with this updated \(\Lambda_{lm}\) to obtain \(\omega^{(1)}\).
Unless stated otherwise, all tabulated and plotted frequencies correspond to \(\omega\equiv \omega^{(1)}\).
Empirically, this one-step update is sufficient to reach the quoted six-digit accuracy for \(a\lesssim 0.9\); near extremality, residual discrepancies reflect slower convergence and root-selection sensitivity.

\paragraph*{Convergence test of the QNM frequencies}
For each parameter set $(a,r_2,\mu,l,m)$, the convergence of the quasinormal-mode frequency $\omega$ obtained from Eq.~\eqref{Eq:characteristic_AIM} was examined by varying the truncation order $N_{\max}$ in the range $9 \leq N_{\max} \leq 49$. The convergence pattern depends systematically on the black-hole spin parameter $a$.

\begin{itemize}
\item \textbf{Moderate rotation} ($a \leq 0.5$):  
  The AIM sequence converges monotonically, and the QNM frequency stabilizes numerically for $N_{\max}\gtrsim 25$.

\item \textbf{Rapid rotation} ($0.5 < a \leq 0.95$):  
  The frequency exhibits a stable plateau in the interval $17\lesssim N_{\max}\lesssim 27$. Outside this range the values progressively deviate.

\item \textbf{Near-extremal rotation} ($a > 0.95$):  
  The sequence converges monotonically up to a stable region around $N_{\max}\approx 15$--$21$, beyond which numerical instability in the recursion coefficients causes the frequency to depart from the plateau.
\end{itemize}

Accordingly, the value of $N_{\max}$ reported in Tables~\ref{tab:qnm_comparison_l1m1}--\ref{tab:qnm_comparison_emda}  is chosen according to the following explicit criterion: we identify the stable region in which $\omega$ changes by less than $5\times10^{-5}$ in $\operatorname{Re}(\omega)$ and less than $5\times10^{-6}$ in $\operatorname{Im}(\omega)$ under the variation $N_{\max}\to N_{\max}\pm1,\pm2$. The adopted $N_{\max}$ (typically 19--37) is then selected inside this interval, guaranteeing six-decimal-place accuracy of the reported AIM frequencies.

\subsection{Results}
Tables~\ref{tab:qnm_comparison_l1m1} and \ref{tab:qnm_comparison_l2m2} present the fundamental-mode (\(n=0\)) QNM frequencies for scalar perturbations on a Kerr black hole, computed using the asymptotic iteration method (AIM) with the two-pass update of the angular separation constant, compared against continued-fraction method (CFM) \cite{Leaver1985} results from Konoplya \& Zhidenko~\cite{Konoplya2006} for \(l=1, m=1\) and Dolan~\cite{Dolan2007} for \(l=2, m=2\). Table~\ref{tab:qnm_comparison_emda} compares AIM results for the Kerr–EMDA spacetime (\(r_2=0.2\) and \(0.4\)) with CFM data from Pan \& Jing~\cite{Pan2006} for \(l=0, m=0\), \(l=2, m=0\), and \(l=2, m=2\). We report \(\operatorname{Re}(\omega)\) and \(-\operatorname{Im}(\omega)\) (positive for damped QNMs) in units of \(M=1\), rounded to six decimal places. Absolute differences, \(\Delta \operatorname{Re}(\omega)\) and \(\Delta(-\operatorname{Im}(\omega))\), quantify the agreement.

For the pure Kerr case, deviations for \(l=1, m=1\) remain below \(1.1\times 10^{-4}\) in \(\operatorname{Re}(\omega)\) and below \(1.0\times 10^{-4}\) in \(-\operatorname{Im}(\omega)\) for \(a\le 0.9\). As \(a\to 1\), discrepancies increase (e.g., at \(a=0.995\) they reach \(\sim 4.3\times 10^{-4}\) in \(\operatorname{Re}(\omega)\) and \(\sim 4.0\times 10^{-4}\) in \(-\operatorname{Im}(\omega)\)), reflecting the known convergence challenges near extremality. For \(l=2, m=2\), the agreement is tighter across the scan, with differences typically below \(1.3\times 10^{-4}\) (\(\operatorname{Re}(\omega)\)) and \(1.7\times 10^{-4}\) (\(-\operatorname{Im}(\omega)\)).

In the Kerr–EMDA comparison for \(r_2=0.2\) and \(0.4\)~(Table~\ref{tab:qnm_comparison_emda}), discrepancies are especially large for \(l=0\) modes at moderate spin, reaching \(\sim 1.1\times 10^{-3}\) in \(\operatorname{Re}(\omega)\) and \(\sim 1.2\times 10^{-3}\) in \(-\operatorname{Im}(\omega)\). These trends can be attributed to the dilaton's modification of the effective potential, raising its barrier height and shifting it inward, which enhances wave confinement and increases sensitivity particularly for low-multipole modes.
\begin{longtable}{ccccccccc}
\caption{Comparison of scalar perturbations on a Kerr black hole for the fundamental mode of quasinormal mode (QNM) frequencies with \(l=1, m=1\). The columns show \(\operatorname{Re}(\omega)\) and \(-\operatorname{Im}(\omega)\) from Konoplya \& Zhidenko (CFM) \cite{Konoplya2006} for \(a\) from 0.0 to 0.99, from this work (CFM) for \(a=0.995\), and from this work (AIM), along with absolute differences.} \label{tab:qnm_comparison_l1m1} \\
\toprule
& & \multicolumn{2}{c}{Konoplya \& Zhidenko} & \multicolumn{3}{c}{This Work (AIM)} & \multicolumn{2}{c}{Difference} \\
\(a\) & \(\mu\)& \(\operatorname{Re}(\omega)\) & \(-\operatorname{Im}(\omega)\) &\( N_{\mathrm{max}}\) & \(\operatorname{Re}(\omega)\) & \(-\operatorname{Im}(\omega)\) & \(\Delta \operatorname{Re}(\omega)\) & \(\Delta (-\operatorname{Im}(\omega))\) \\
\midrule
\endfirsthead
\toprule
& & \multicolumn{2}{c}{Konoplya \& Zhidenko} & \multicolumn{3}{c}{This Work (AIM)} & \multicolumn{2}{c}{Difference} \\
\(a\) & \(\mu\) & \(\operatorname{Re}(\omega)\) & \(-\operatorname{Im}(\omega)\) &\( N_{\mathrm{max}}\) &\(\operatorname{Re}(\omega)\) & \(-\operatorname{Im}(\omega)\) & \(\Delta \operatorname{Re}(\omega)\) & \(\Delta (-\operatorname{Im}(\omega))\) \\
\midrule
\endhead
\multicolumn{9}{r}{{\emph{Continued on next page}}} \\
\midrule
\endfoot
\bottomrule
\endlastfoot
0.0 & 0.0 & 0.292936 & 0.097660 & 37 & 0.292941 & 0.097660 & 0.000005 & 0.000000\\
 & 0.1 & 0.297416 & 0.094957 & 37 & 0.297422 & 0.094957 & 0.000006 & 0.000000 \\
 & 0.2 & 0.310957 & 0.086593 & 37 & 0.310974 & 0.086592 & 0.000017 & 0.000001 \\
0.1 & 0.0 & 0.301045 & 0.097547 & 37 & 0.301049 & 0.097545 & 0.000004 & 0.000002 \\
 & 0.1 & 0.305329 & 0.095029 & 37 & 0.305335 & 0.095026 & 0.000006 & 0.000003 \\
 & 0.2 & 0.318274 & 0.087228 & 37 & 0.318291 & 0.087223 & 0.000017 & 0.000005\\
0.2 & 0.0 & 0.310043 & 0.097245 & 27 & 0.310043 & 0.097264 & 0.000000 & 0.000019 \\
 & 0.1 & 0.314119 & 0.094920 & 27 & 0.314116 & 0.094942 & 0.000003 & 0.000022\\
 & 0.2 & 0.326433 & 0.087709 & 27 & 0.326429 & 0.087760 & 0.000004 & 0.000051\\
0.3 & 0.0 & 0.320126 & 0.096691 & 37 & 0.320129 & 0.096689 & 0.000003 & 0.000002 \\
 & 0.1 & 0.323981 & 0.094569 & 37 & 0.323984 & 0.094566 & 0.000003 & 0.000003\\
 & 0.2 & 0.335621 & 0.087979 & 37 & 0.335630 & 0.087974 & 0.000009 & 0.000005\\
0.4 & 0.0 & 0.331567 & 0.095792 & 37 & 0.331570 & 0.095789 & 0.000003 & 0.000003 \\
 & 0.1 & 0.335181 & 0.093883 & 37 & 0.335186 & 0.093881 & 0.000005 & 0.000002\\
 & 0.2 & 0.346095 & 0.087950 & 37 & 0.346107 & 0.087946 & 0.000012 & 0.000004 \\
0.5 & 0.0 & 0.344753 & 0.094395 & 37 & 0.344758 & 0.094392 & 0.000005 & 0.000003 \\
 & 0.1 & 0.348105 & 0.092714 & 37 & 0.348111 & 0.092711 & 0.000006 & 0.000003\\
 & 0.2 & 0.358230 & 0.087478 & 37 & 0.358242 & 0.087476 & 0.000012 & 0.000002 \\
0.6 & 0.0 & 0.360285 & 0.092243 & 23 & 0.360260 & 0.092262 & 0.000025 & 0.000019 \\
 & 0.1 & 0.363345 & 0.090805 & 23 & 0.363316 & 0.090829 & 0.000029 & 0.000024\\
 & 0.2 & 0.372594 & 0.086320 & 23 & 0.372539 & 0.086363 & 0.000055 & 0.000043 \\
0.7 & 0.0 & 0.379159 & 0.088848 & 23 & 0.379113 & 0.088861 & 0.000046 & 0.000013 \\
 & 0.1 & 0.381888 & 0.087678 & 23 & 0.381834 & 0.087691 & 0.000054 & 0.000013 \\
 & 0.2 & 0.390141 & 0.084014 & 23 & 0.390054 & 0.084031 & 0.000087 & 0.000017 \\
0.8 & 0.0 & 0.403273 & 0.083132 & 23 & 0.403211 & 0.083125 & 0.000062 & 0.000007 \\
 & 0.1 & 0.405606 & 0.082262 & 23 & 0.405537 & 0.082252 & 0.000069 & 0.000010 \\
 & 0.2 & 0.412675 & 0.079526 & 23 & 0.412570 & 0.079505 & 0.000105 & 0.000021 \\
0.9 & 0.0 & 0.437234 & 0.071848 & 23 & 0.437163 & 0.071794 & 0.000071 & 0.000054 \\
 & 0.1 & 0.439045 & 0.071342 & 23 & 0.438968 & 0.071279 & 0.000077 & 0.000063 \\
 & 0.2 & 0.444549 & 0.069737 & 23 & 0.444449 & 0.069636 & 0.000100 & 0.000101 \\
0.95 & 0.0 & 0.462261 & 0.060091 & 23 & 0.462288 & 0.059941 & 0.000027 & 0.000150 \\
 & 0.1 & 0.463691 & 0.059825 & 23 & 0.463727 & 0.059659 & 0.000036 & 0.000166 \\
 & 0.2 & 0.468050 & 0.058968 & 23 & 0.468135 & 0.058751 & 0.000085 & 0.000217 \\
0.99 & 0.0 & 0.493423 & 0.036712 & 19 & 0.493492 & 0.037240 & 0.000069 & 0.000528 \\
 & 0.1 & 0.494284 & 0.036756 & 19 & 0.494297 & 0.037324 & 0.000013 & 0.000568 \\
 & 0.2 & 0.496939 & 0.036879 & 19 & 0.496931 & 0.037635 & 0.000008 & 0.000756 \\
 0.995 & 0.0 & 0.499105 & 0.029271 & 19 & 0.498677 & 0.029532 & 0.000428 & 0.000261 \\
 & 0.1 & 0.499800 & 0.029394 & 19 & 0.499438 & 0.029774 & 0.000362 & 0.000380 \\
 & 0.2 & 0.501956 & 0.029776 & 19 & 0.501328 & 0.030082 & 0.000628 & 0.000306 \\
\end{longtable}
\begin{longtable}{ccccccccc}
\caption{Comparison of scalar perturbations on Kerr black hole for fundamental mode of QNM frequencies \(l=2, m=2\) from Dolan (CFM) \cite{Dolan2007} and this work (AIM). The numbers with asterisk are from Pan \& Jing (CFM) \cite{Pan2006}. All values are in units of black hole mass \(M=1\).} \label{tab:qnm_comparison_l2m2} \\
\toprule
&& \multicolumn{2}{c}{Dolan (CFM)} & \multicolumn{3}{c}{This Work (AIM)} & \multicolumn{2}{c}{Difference} \\
\(a\) & \(\mu\) & \(\operatorname{Re}(\omega)\) & \(-\operatorname{Im}(\omega)\) &\( N_{\mathrm{max}}\) & \(\operatorname{Re}(\omega)\) & \(-\operatorname{Im}(\omega)\) & \(\Delta \operatorname{Re}(\omega)\) & \(\Delta (-\operatorname{Im}(\omega))\) \\
\midrule
\endfirsthead
\toprule
& & \multicolumn{2}{c}{Dolan (CFM)} & \multicolumn{3}{c}{This Work (AIM)} & \multicolumn{2}{c}{Difference} \\
\(a\) & \(\mu\) & \(\operatorname{Re}(\omega)\) & \(-\operatorname{Im}(\omega)\) &\( N_{\mathrm{max}}\) & \(\operatorname{Re}(\omega)\) & \(-\operatorname{Im}(\omega)\) & \(\Delta \operatorname{Re}(\omega)\) & \(\Delta (-\operatorname{Im}(\omega))\) \\
\midrule
\endhead
\multicolumn{9}{r}{{\emph{Continued on next page}}} \\
\midrule
\endfoot
\bottomrule
\endlastfoot
0.0 & 0.0 & 0.483644 & 0.096759 & 31 & 0.483644 & 0.096758 & 0.000000 & 0.000001 \\
 & \(0.0^{*}\) & \(0.483644^{*}\) & \(0.096759^{*}\) & & & & & \\
 & 0.1 & 0.486804 & 0.095675 & 31&0.486804 & 0.095674 & 0.000000 & 0.000001 \\
 & 0.2 & 0.496327 & 0.092389 & 31& 0.496327 & 0.092388 & 0.000000 & 0.000001 \\
 & 0.3 & 0.512346 & 0.086795 & 31 & 0.512348 & 0.086793 & 0.000002 & 0.000002 \\
0.1 & 0.0 & 0.499482 & 0.096666 & 31 & 0.499482& 0.096666 & 0.000000 & 0.000000 \\
 & 0.1 & 0.502456 & 0.095674 & 30 &0.502456 & 0.095673 & 0.000000 & 0.000001 \\
 & 0.2 & 0.511419 & 0.092663 & 31 & 0.511419 & 0.092663 & 0.000000 & 0.000000 \\
 & 0.3 & 0.526497 & 0.087528 & 31 & 0.526498& 0.087527 & 0.000001 & 0.000001 \\
0.2 & 0.0 & 0.517121 & 0.096382& 31 &0.517116 & 0.096369 & 0.000005 & 0.000013 \\
 & \(0.0^{*}\) & \(0.517120^{*}\) & \(0.096382^{*}\) & & & & & \\
 & 0.1 & 0.519901 & 0.095483 & 31 &0.519902 & 0.095476 & 0.000001 & 0.000007 \\
 & 0.2 & 0.528281 & 0.092755 & 31& 0.528284 & 0.092748 & 0.000003 & 0.000007 \\
 & 0.3 & 0.542378 & 0.088092 & 31 & 0.542391 & 0.088070 & 0.000013 & 0.000022 \\
0.3 & 0.0 & 0.536979 & 0.095839 &31& 0.536979 & 0.095838 & 0.000000 & 0.000001 \\
 & 0.1 & 0.539557 & 0.095036 & 31 & 0.539557 & 0.095035 & 0.000000 & 0.000001 \\
 & 0.2 & 0.547326 & 0.092595 & 31 & 0.547326 & 0.092595 & 0.000000 & 0.000000 \\
 & 0.3 & 0.560397 & 0.088418 & 31 & 0.560397 & 0.088417 & 0.000000 & 0.000001 \\
0.4 & 0.0 & 0.559647 & 0.094931 & 31 & 0.559646 & 0.094931 & 0.000001 & 0.000000 \\
 & \(0.0^{*}\) & \(0.559645^{*}\) & \(0.094931^{*}\) & & & & & \\
 & 0.1 & 0.562011 & 0.094226 & 31& 0.562010 & 0.094226& 0.000001 & 0.000000 \\
 & 0.2 & 0.569137 & 0.092082 & 31& 0.569136 & 0.092082 & 0.000001 & 0.000000 \\
 & 0.3 & 0.581127 & 0.088406 & 31 & 0.581126 & 0.088405 & 0.000001 & 0.000001 \\
0.5 & 0.0 & 0.585990 & 0.093494 & 31 & 0.585989 & 0.093494 & 0.000001 & 0.000000 \\
 & 0.1 & 0.588127 & 0.092890 & 31 & 0.588125 & 0.092890 & 0.000002 & 0.000000 \\
 & 0.2 & 0.594568 & 0.091052 & 31 & 0.594567 & 0.091052 & 0.000001 & 0.000000 \\
 & 0.3 & 0.605408 & 0.087895 & 31 &0.605406& 0.087894& 0.000002 & 0.000001 \\
0.6 & 0.0 & 0.617364 & 0.091245 & 25 & 0.617363 & 0.091243 & 0.000001 & 0.000002 \\
 & \(0.0^{*}\) & \(0.617365^{*}\) & \(0.091246^{*}\) & & & & & \\
 & 0.1 & 0.619256 & 0.090746 & 25 & 0.619255 & 0.090744 & 0.000001 & 0.000002 \\
 & 0.2 & 0.624959 & 0.089224 & 25 & 0.624959 & 0.089222 & 0.000000 & 0.000002 \\
 & 0.3 & 0.634559 & 0.086607& 25 &0.634561 & 0.086604& 0.000002 & 0.000003 \\
0.7 & 0.0 & 0.656099 & 0.087649 & 25 & 0.656100 & 0.087648 & 0.000001 & 0.000001 \\
 & 0.1 & 0.657722 & 0.087259 & 24 & 0.657719 & 0.087255 & 0.000003 & 0.000004 \\
 & 0.2 & 0.662614 & 0.086068 & 25 & 0.662616 & 0.086067 & 0.000002 & 0.000001 \\
 & 0.3 & 0.670848 & 0.084018 & 25 & 0.670853 & 0.084018 & 0.000005 & 0.000000 \\
0.8 & 0.0 & 0.706823 & 0.081520 & 25 &0.706820 & 0.081518 & 0.000003 & 0.000002 \\
 & \(0.0^{*}\) & \(0.706825^{*}\) & \(0.081520^{*}\)& & & & & \\
 & 0.1 & 0.708138 & 0.081245 & 25& 0.708135 & 0.081244 & 0.000003 & 0.000001 \\
 & 0.2 & 0.712102 & 0.080407 & 25 & 0.712100 & 0.080406 & 0.000002 & 0.000001 \\
 & 0.3 & 0.718777 & 0.078961 & 24 & 0.718775& 0.078952 & 0.000002 & 0.000009 \\
0.9 & 0.0 & 0.781638 & 0.069289 & 25 & 0.781629 & 0.069283 & 0.000009 & 0.000006 \\
 & \(0.0^{*}\) & \(0.781640^{*}\) & \(0.069289^{*}\)& & & & & \\
 & 0.1 & 0.782570 & 0.069142 & 25 & 0.782561 & 0.069135& 0.000009 & 0.000007 \\
 & 0.2 & 0.785380 & 0.068690 & 24 & 0.785374 & 0.068686 & 0.000006 & 0.000004 \\
 & 0.3 & 0.790113 & 0.067911 & 24 & 0.790111 & 0.067907 & 0.000002 & 0.000004 \\
0.95 & 0.0 & 0.840982 & 0.056471 & 25 & 0.840967 & 0.056479 & 0.000015 & 0.000008 \\
 & 0.1 & 0.841653 & 0.056395 & 25& 0.841637 & 0.056403 & 0.000016 & 0.000008 \\
 & 0.2 & 0.843677 & 0.056163 & 24 & 0.843661 & 0.056173 & 0.000016 & 0.000010 \\
 & 0.3 & 0.847088 & 0.055762 & 25 & 0.847071 & 0.055774 & 0.000017 & 0.000012 \\
0.99 & 0.0 & 0.928028 & 0.031063 & 19 & 0.927900 & 0.030960 & 0.000128 & 0.000103 \\
 & 0.1 & 0.928353 & 0.031054 & 19& 0.928253 & 0.030959 & 0.000100 & 0.000095 \\
 & 0.2 & 0.929336 & 0.031026 & 19 & 0.929207 & 0.030901 & 0.000129 & 0.000125 \\
 & 0.3 & 0.930994 & 0.030977 & 19& 0.930868& 0.030817 & 0.000126 & 0.000160 \\
0.995 & 0.0 & 0.949522 & 0.023104 & 19& 0.949492 & 0.022949 & 0.000030 & 0.000155 \\
 & 0.1 & 0.949762 & 0.023104 & 19 & 0.949774 & 0.023014 & 0.000012 & 0.000090 \\
 & 0.2 & 0.950487 & 0.023102 & 19 & 0.950493 & 0.022973 & 0.000006 & 0.000129 \\
 & 0.3 & 0.951712 & 0.023100 & 19 & 0.951726 & 0.022936 & 0.000014 & 0.000164 \\
\end{longtable}
\begin{longtable}{cccccccccc}
\caption{Comparison of scalar perturbations on Kerr-EMDA black hole with dilaton parameters \(r_2=0.2\) amd \(0.4\) for fundamental mode of QNM frequencies from Pan \& Jing (CFM) \cite{Pan2006} and this work (AIM). All values are calculated in units of black hole mass \(M=1\).} \label{tab:qnm_comparison_emda} \\
\toprule
& & & \multicolumn{2}{c}{Pan \& Jing (CFM)} & \multicolumn{3}{c}{This Work (AIM)} & \multicolumn{2}{c}{Difference} \\
\(a\) & \(r_2\)  & \(\mu\) & \(\operatorname{Re}(\omega)\) & \(-\operatorname{Im}(\omega)\) &\( N_{\mathrm{max}}\) & \(\operatorname{Re}(\omega)\) & \(-\operatorname{Im}(\omega)\) & \(\Delta \operatorname{Re}(\omega)\) & \(\Delta (-\operatorname{Im}(\omega))\) \\
\midrule
\endfirsthead
\toprule
& & & \multicolumn{2}{c}{Pan \& Jing (CFM)} & \multicolumn{3}{c}{This Work (AIM)} & \multicolumn{2}{c}{Difference} \\
\(a\) & \(r_2\)  & \(\mu\) & \(\operatorname{Re}(\omega)\) & \(-\operatorname{Im}(\omega)\) &\(N_{\mathrm{max}}\) & \(\operatorname{Re}(\omega)\) & \(-\operatorname{Im}(\omega)\) & \(\Delta \operatorname{Re}(\omega)\) & \(\Delta (-\operatorname{Im}(\omega))\) \\
\midrule
\endhead
\multicolumn{10}{r}{{\emph{Continued on next page}}} \\
\midrule
\endfoot
\bottomrule
\endlastfoot
 \multicolumn{10}{c}{\(l=0,\ m=0\)} \\
0.00 & 0.2 & 0.0 & 0.119106 & 0.106684 & 37 & 0.118490 & 0.106336 & 0.000616 & 0.000348 \\
0.20 & 0.2 & 0.0 & 0.119539 & 0.106113 & 37 & 0.118681 & 0.105696 & 0.000858 & 0.000417 \\
0.40 & 0.2 & 0.0 & 0.120787 & 0.104129 & 37 & 0.120485 & 0.103171 & 0.000302 & 0.000958 \\
0.60 & 0.2 & 0.0 & 0.122300 & 0.099555 & 37 & 0.122858 & 0.098405 & 0.000558 & 0.001150 \\
 \multicolumn{10}{c}{\(l=2,\ m=0\)} \\
\midrule
0.00 & 0.2 & 0.0 & 0.520175 & 0.099005 & 37 & 0.520177 & 0.099005 & 0.000002 & 0.000000 \\
0.20 & 0.2 & 0.0 & 0.522050 & 0.098509 & 37 & 0.522051 & 0.098508 & 0.000001 & 0.000001 \\
0.40 & 0.2 & 0.0 & 0.527965 & 0.096821 & 37 & 0.527962 & 0.096819 & 0.000003 & 0.000002 \\
0.60 & 0.2 & 0.0 & 0.538950 & 0.093080 & 37 & 0.538938 & 0.093077 & 0.000012 & 0.000003 \\
 \multicolumn{10}{c}{\(l=2,\ m=2\)} \\
\midrule
0.00 & 0.2 &0.0 & 0.520175 & 0.099005 & 37 &0.520177 & 0.099005 & 0.000002 & 0.000000 \\
0.20 & 0.2 & 0.0 & 0.562935 & 0.098011 & 37 & 0.562934 & 0.098011 & 0.000001 & 0.000000 \\
0.40 & 0.2 & 0.0 & 0.621050 & 0.094891 & 25 & 0.621057 & 0.094899 & 0.000007 & 0.000008 \\
0.60 & 0.2 & 0.0 & 0.711290 & 0.085877 & 25 &0.711291 & 0.085884 & 0.000001 & 0.000007 \\
\midrule
 \multicolumn{10}{c}{\(l=0,\ m=0\)} \\
0.00 & 0.4 & 0.0 & 0.130293 & 0.108633 & 37 & 0.130221 & 0.107652 & 0.000072 & 0.000981 \\
0.20 & 0.4 & 0.0 & 0.130895& 0.107528 & 37 &  0131445& 0.107896& 0.000550 & 0.000368 \\
0.40 & 0.4 & 0.0 & 0.132289 & 0.103312 & 37 & 0.133344 & 0.102457 & 0.001055 & 0.000855 \\
 \multicolumn{10}{c}{\(l=2,\ m=0\)} \\
\midrule
0.00 & 0.4 & 0.0 & 0.568130 & 0.101550 & 37 & 0.568130 & 0.101550 & 0.000000 & 0.000000 \\
0.20 & 0.4 & 0.0 & 0.571195 & 0.100626 & 37 & 0.571195 & 0.100624& 0.000000 & 0.000002 \\
0.40 & 0.4 & 0.0 & 0.581175& 0.097196 & 37 & 0.581170 & 0.097201& 0.000005 & 0.000005 \\
 \multicolumn{10}{c}{\(l=2,\ m=2\)} \\
\midrule
0.00 & 0.4 & 0.0 & 0.568130 & 0.101550 & 37 & 0.568130 & 0.101550 & 0.000000 & 0.000000 \\
0.20 & 0.4 & 0.0 & 0.627000 & 0.099226 & 37 & 0.627000 & 0.099224& 0.000000 & 0.000002 \\
0.40 & 0.4 & 0.0 & 0.718640 & 0.091296 & 25 & 0.718623& 0.091298& 0.000017 & 0.000002 \\
\end{longtable}

\subsubsection{Validation Against Konoplya \& Zhidenko (2006)}
\begin{figure}
\centering
\includegraphics[width=0.8\textwidth]{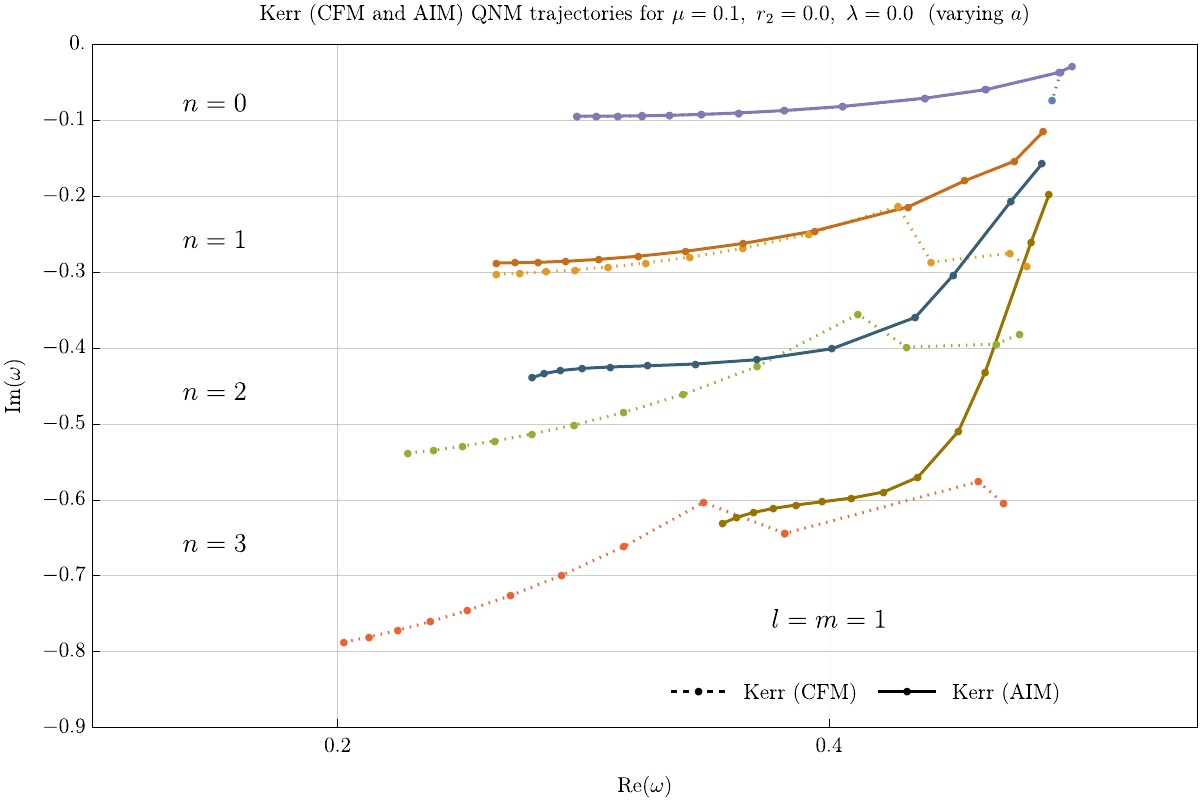}
\caption{Comparison of QNM trajectories for \(l=1, m=1\) between AIM (solid lines) and Konoplya \& Zhidenko (2006) (dashed lines) for Kerr black hole with \(\mu = 0.1\) (varying \(a\)).}
\label{fig:kerr_comparison}
\end{figure}

Figure~\ref{fig:kerr_comparison} shows QNM trajectories in the complex \(\omega\)-plane for \(l=1, m=1\), \(\mu=0.1\), and overtones \(n=0\)–3, comparing AIM (solid) with the CFM results of Konoplya \& Zhidenko~\cite{Konoplya2006} (dashed) for \(0\le a\le 0.995\). For the fundamental mode \(n=0\), absolute deviations remain below \(1.0\times 10^{-4}\) in \(\operatorname{Re}(\omega)\) and below \(2.2\times 10^{-5}\) in \(-\operatorname{Im}(\omega)\) for \(a\leq 0.95\). Near extremality (\(a=0.99\) and \(0.995\)) they increase to \(\sim 6.3\times 10^{-4}\) in \(\operatorname{Re}(\omega)\) and \(\sim 7.6\times 10^{-4}\) in \(-\operatorname{Im}(\omega)\), consistent with the slower convergence of semi-analytic schemes and heightened root-selection sensitivity in the near-extremal regime. Overall, the trajectories corroborate the reliability of AIM at moderate spin while delineating its limitations as \(a\to 1\).

\subsubsection{Comparison with WKB Trends}
The present AIM results follow the qualitative trends established in earlier WKB studies of Kerr QNMs~\cite{Seidel1990,Simone1992}. As \(a\) increases, prograde modes exhibit increasing \(\mathrm{Re}(\omega)\) and decreasing \(|\mathrm{Im}(\omega)|\), reflecting the rotational “spin-up” effect: higher oscillation frequencies and slower damping. For massive scalar fields (\(\mu > 0\)), as in Tables~\ref{tab:qnm_comparison_l1m1} and \ref{tab:qnm_comparison_l2m2}, the trajectories are shifted relative to the massless case, with larger \(\mu\) yielding higher \(\mathrm{Re}(\omega)\) and weaker damping. This behaviour, seen in Simone \& Will~\cite{Simone1992}, arises from modifications to the effective potential induced by the mass term. Our agreement with the CFM data confirms that AIM captures these rotational and mass effects with fidelity for low-lying modes.

\subsubsection{Effect of Dilaton Parameter \(r_2\)}
Figure~\ref{fig:kerr_vs_emda} compares pure Kerr (\(r_2=0\), dotted) with Kerr–EMDA (\(r_2=0.1\), solid) at \(\lambda=0\). For both \(l=m=1\) and \(l=m=2\), overtones \(n=0\)–3 shift rightward (larger \(\mathrm{Re}(\omega)\)) and downward (more negative \(\mathrm{Im}(\omega)\)) with increasing \(r_2\). This indicates higher oscillation frequencies but faster decay. In EMDA theory, the dilaton couples to the electromagnetic sector, modifying the near-horizon geometry and enhancing energy dissipation. Similar down–right shifts have been reported in other dilaton black hole models~\cite{Pan2006,Pacilio2018,Malik2024}.

\subsubsection{Effect of PFDM Parameter \(\lambda\)}
\begin{figure}
\centering
\includegraphics[width=0.8\textwidth]{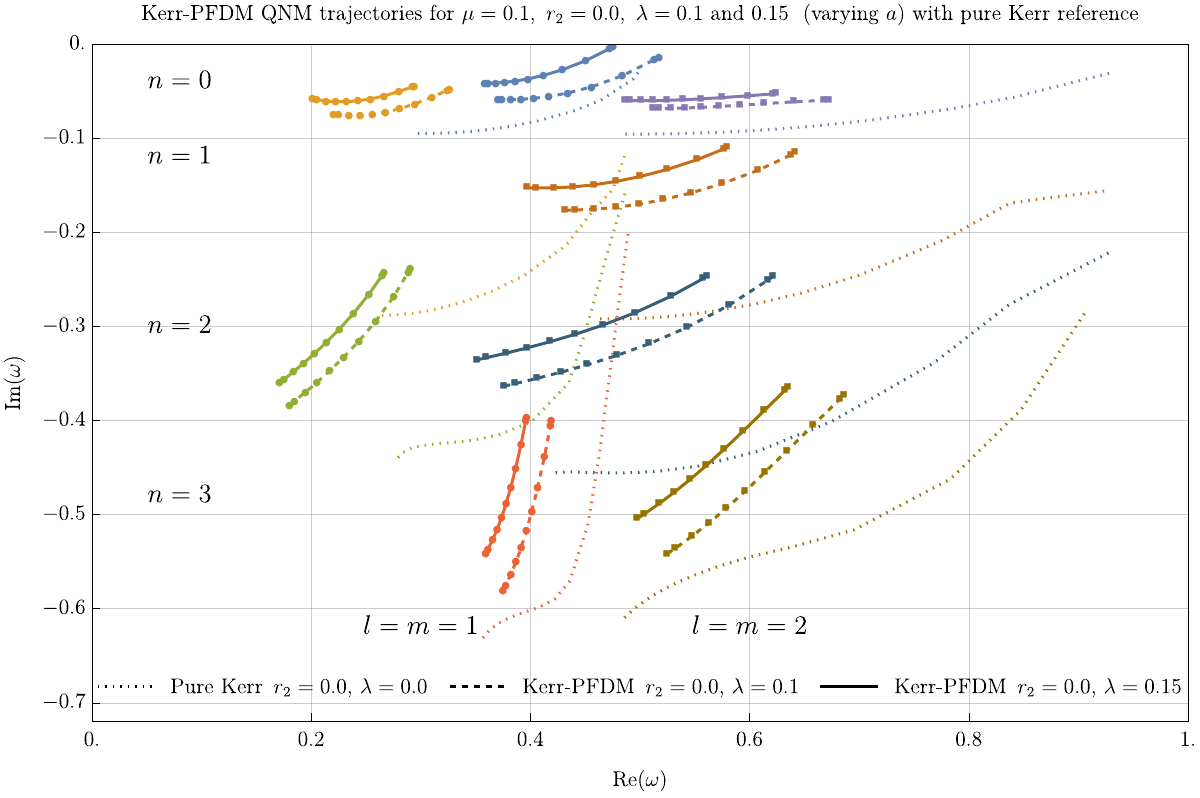}
\caption{QNM trajectories for Kerr–PFDM (\(r_2=0.0\), \(\lambda = 0.1, 0.15\)) with \(\mu = 0.1\), compared with pure Kerr. Increasing \(\lambda\) lowers the frequency and reduces damping. The points shown are for the values $a=0.575, 0.6, 0.65, 0.7, 0.75, 0.8, 0.85, 0.9, 0.95, 0.99$ and $ 0.995$.}
\label{fig:kerr_vs_pfdm}
\end{figure}

Figure~\ref{fig:kerr_vs_pfdm} compares pure Kerr (\(\lambda=0\), dotted) with Kerr–PFDM at \(\lambda=0.1\) (dashed) and \(\lambda=0.15\) (solid), keeping \(r_2=0\). Increasing \(\lambda\) produces a leftward (smaller \(\mathrm{Re}(\omega)\)) and upward (less negative \(\mathrm{Im}(\omega)\)) shift, implying reduced oscillation frequencies but longer-lived modes. This effect is more pronounced than for \(r_2\). In the PFDM model, the logarithmic potential term due to the dark matter halo “softens” the effective barrier for perturbations, suppressing frequencies while reducing damping, consistent with analyses in~\cite{Sun2023,Das2024,Wang2024,Tan2025,Jusufi2020}.

\subsubsection{Combined Effects: Kerr–EMDA–PFDM}
\begin{figure}
\centering
\includegraphics[width=0.8\textwidth]{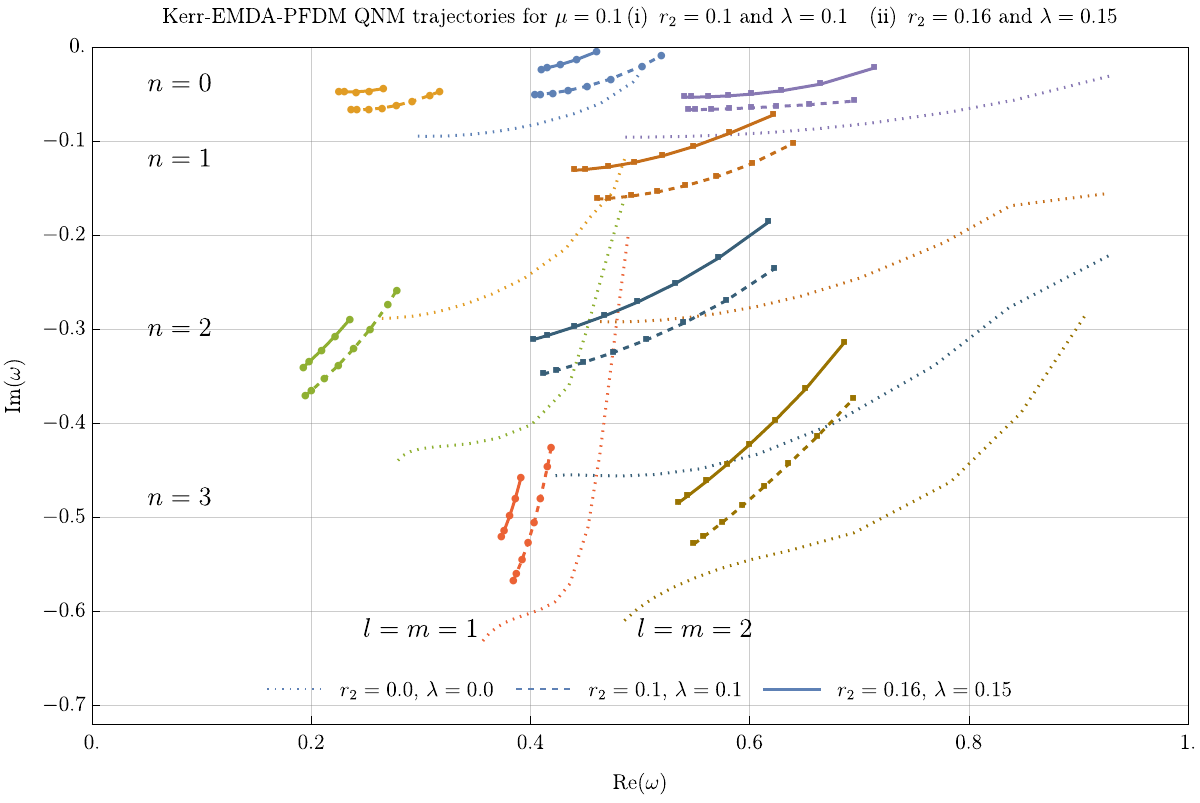}
\caption{Combined effects of \(r_2\) and \(\lambda\) on QNM trajectories for \(\mu = 0.1\). PFDM’s up–left shift partially offsets the dilaton’s down–right trend. The points shown are for the values $a=0.575, 0.6, 0.65, 0.7, 0.75, 0.8, 0.85$ and $ 0.9$.}
\label{fig:emda_pfdm}
\end{figure}

Figure~\ref{fig:emda_pfdm} illustrates the combined influence of \(r_2\) and \(\lambda\), for \(r_2=0.1, \lambda=0.1\) (dashed) and \(r_2=0.16, \lambda=0.15\) (solid), compared with pure Kerr (dotted). The dilaton term alone drives a down–right shift, but the PFDM term dominates with its up–left shift, resulting in net trajectories that lie above and slightly to the left of the Kerr baseline. The nonlinear interplay suggests that PFDM’s stabilizing influence can partially counteract dilaton-induced damping, potentially indicating parameter regimes with enhanced mode stability. While the present analysis is in the linear regime, these qualitative trends provide guidance for future time-domain and stability studies.

\begin{figure}
\centering
\includegraphics[width=0.8\textwidth]{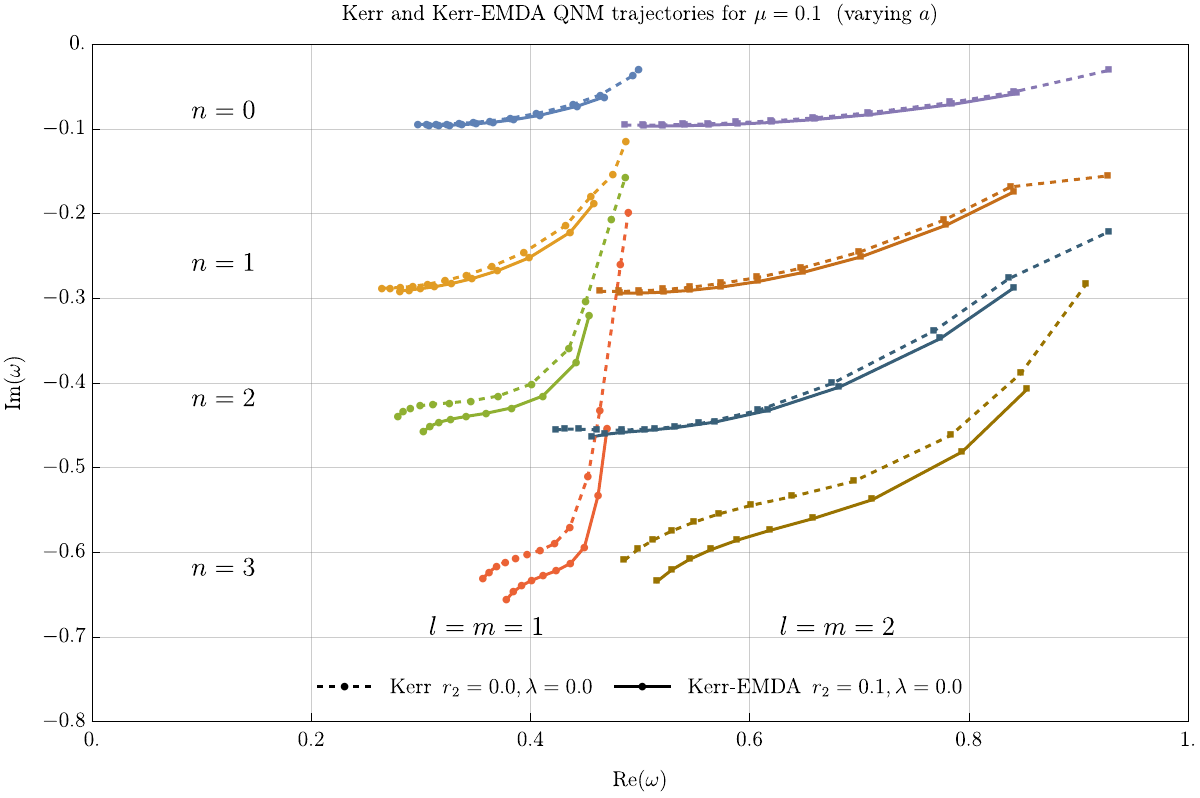}
\caption{QNM trajectories for Kerr (\(r_2=0\)) and Kerr–EMDA (\(r_2=0.1\)) with \(\mu = 0.1\) and varying \(a\). Higher spins shift modes to higher frequencies and increase damping.}
\label{fig:kerr_vs_emda}
\end{figure}

\section{Discussion}

We have investigated the dynamics of a massive scalar field in the background of a Kerr black hole within the Einstein–Maxwell–dilaton–axion (EMDA) framework, embedded in a perfect fluid dark matter (PFDM) halo. Our analysis has focused on superradiance and quasinormal modes (QNMs), highlighting the contrasting roles of the dilaton parameter \( r_2 \) and the PFDM parameter \( \lambda \) in shaping the spacetime geometry, effective potential, and perturbation spectra.

The EMDA–PFDM spacetime, described by Eq.~\eqref{Eq:EMDA-PFDM_metric} with \(\Delta_D(r)\) in Eq.~\eqref{Def:rho2_DeltaD}, exhibits distinct horizons and ergosphere structures depending on \(r_2\) and \(\lambda\) (Fig.~\ref{fig:deltaplots}). Increasing \(r_2\) (for fixed \(\lambda\)) reduces the event horizon radius \(r_h\) relative to the infinite redshift surface \(r_{rs}\), resulting in a larger ergosphere. This behaviour arises from the dilaton’s coupling to electromagnetic and axion fields, which reduces the frame-dragging region~\cite{Ganguly2016,Banerjee2020}. In contrast, increasing \(\lambda\) expands \(r_h\) relative to \(r_{rs}\), shrinking the ergosphere through the PFDM’s logarithmic gravitational contribution~\cite{Li2012,Das2021,Liang2023,Mollicone2025}. These geometric changes modify the horizon angular velocity \(\Omega_h\) and surface gravity, thereby influencing the superradiance condition and the damping properties of perturbations.

The effective potential \(V_{\mathrm{eff}}(r)\) in Eq.~\eqref{Eq:Veff} is likewise sensitive to \(r_2\) and \(\lambda\) (Figs.~\ref{fig:PotentialPlotsRealOmega} and \ref{fig:PotentialPlotsComplexOmega}). For fixed \(\lambda\), larger \(r_2\) raises the barrier height and shifts it inward, enhancing wave confinement and scattering efficiency. For fixed \(r_2\), larger \(\lambda\) lowers the barrier and shifts it outward, effectively softening the potential. These modifications are reflected in the superradiance amplification factor \(Z_{lm}\) (Eq.~\eqref{Eq:Zlm-EMDA-PDFM}), computed via asymptotic matching in Sec.~\ref{Sec:Superradiance-Asymptotic_Matching_Method}. As shown in Figs.~\ref{fig:ZlmSpin} and \ref{fig:Zlm}, \(Z_{lm}\) increases with \(a\) and \(r_2\) (extending the frequency range satisfying \(\omega < m\Omega_h\)), but decreases with \(\lambda\) (shortening that range), in agreement with the corresponding changes in ergosphere size and with results for pure EMDA black holes~\cite{Senjaya2025}.

The net energy extraction rate \(\dot{E}_{\mathrm{net}}\)(Sec.~\ref{subSec:Energy_Extraction}) follows the same trends. Figures~\ref{NetExtractedEnergy_varying_a} and \ref{NetExtractedEnergy_varying_r2_lambda} and Tables~\ref{Ergosphere_varying_a}–\ref{Ergosphere_varying_r2_lambda} show that increasing \(a\) or \(r_2\) enhances extraction via a larger ergosphere, whereas increasing \(\lambda\) suppresses it. The ergosphere width \(r_{\mathrm{ergo}} = r_{rs} - r_h\) grows with spin, is moderately affected by \(r_2\), and decreases with \(\lambda\), consistent with the amplification trends.

Superradiant instability requires both a trapping mechanism (bound-state condition \(\omega < \mu\)) and superradiant amplification (condition \(\omega < m\Omega_H\)). As discussed in Sec.~\ref{Sec:Superradiant_Instability}, the asymptotic form of the effective potential provides a useful heuristic: for \(\omega \gtrsim \mu/\sqrt{2}\), the slope \(V'_{\rm eff}(r)\) becomes positive at large \(r\), suggesting the possible formation of a local minimum capable of supporting quasi-bound states. The PFDM parameter \(\lambda\) deepens this trapping well through its logarithmic correction, yet our numerical results show that the superradiant amplification factor \(Z_{lm}\) and the extracted rotational energy both decrease with increasing \(\lambda\). In the present EMDA black-hole background with PFDM, the suppression of horizon energy extraction therefore dominates over the enhanced trapping, giving rise to a net stabilizing tendency.

The quasi-bound-state analysis presented in Sec.~\ref{Sec:Quasi-bound_States} (extending Detweiler’s treatment for Kerr~\cite{detweiler1980black}) preserves the hydrogen-like asymptotic structure, now modified by the effective mass \(M-r_2\) and the PFDM-induced logarithmic correction. Numerical solutions near the superradiant threshold confirm that increasing \(r_2\) produces a more spatially localized scalar cloud, while increasing \(\lambda\) yields a more extended configuration. In the limit \(\lambda\to0\), our results recover the pure EMDA spectrum~\cite{Senjaya2025}, demonstrating that finite \(\lambda\) introduces a nontrivial environmental modification to the scalar-cloud profile.

The QNM analysis in Sec.~\ref{sec:QNM-EMDA-PFDM}, performed via the asymptotic iteration method (AIM), confirms these parameter effects. Agreement with CFM benchmarks~\cite{Konoplya2006,Dolan2007,Pan2006} is excellent for the fundamental mode and the first overtone except near extremality, where AIM convergence degrades. Figures~\ref{fig:kerr_vs_emda}–\ref{fig:emda_pfdm} show that \(r_2\) shifts QNM trajectories down–right in the complex \(\omega\)-plane (higher \(\mathrm{Re}(\omega)\), more negative \(\mathrm{Im}(\omega)\)), while \(\lambda\) shifts them up–left (lower \(\mathrm{Re}(\omega)\), less negative \(\mathrm{Im}(\omega)\)), with the PFDM effect being more pronounced. These trends are consistent with WKB predictions~\cite{Seidel1990,Simone1992}, and in combination, \(\lambda\) can partially offset dilaton-induced damping, leading to longer-lived modes.

Relative to the standard Kerr case, the EMDA–PFDM spacetime exhibits qualitatively new behaviour: PFDM suppresses superradiance and instability, analogous to the stabilizing influence of galactic dark matter halos~\cite{Sun2023,Das2024,Wang2024,Jusufi2020}, while the dilaton enhances them~\cite{Pacilio2018,Malik2024}. These opposing effects have potential observational consequences. In dark matter–rich environments, PFDM-induced suppression would weaken gravitational-wave ringdown signals and superradiant growth, while strong dilaton coupling could counteract this trend. Future high-precision ringdown measurements with EHT or LISA imaging~\cite{Akiyama2022,Auclair2023,Pitte2023} could, in principle, constrain \(\lambda\) and \(r_2\) through deviations from Kerr QNM spectra and instability timescales.

\section{Conclusion}
We have analyzed massive scalar perturbations of Kerr–EMDA–PFDM black holes, spanning superradiance, quasi-bound states, and QNMs. The dilaton parameter \(r_2\) tends to enhance superradiance and mode growth by enlarging the ergosphere and increasing the superradiant amplification factor, whereas the PFDM parameter \(\lambda\) counteracts these effects by shrinking the ergosphere and suppressing superradiant amplification. Quasi-bound states display hydrogen-like spectra shifted by PFDM, which weakens binding, thereby extending pure-EMDA results~\cite{Senjaya2025}. Our AIM computations of QNMs, validated against continued-fraction benchmarks, are consistent with this picture and indicate a net stabilizing role of PFDM.

Two directions would substantially complete the EMDA–PFDM program. First, a time-domain treatment of scalar scattering—from ringdown to late-time tails—would pin down the tail exponents and their dependence on \(r_2\) and \(\lambda\). Second, extending to gravitational (and possibly electromagnetic/vector) perturbations is essential; whether a Teukolsky-type separable master equation exists on Kerr–EMDA–PFDM remains open, in which case a 2+1D time-domain evolution in horizon-regular coordinates is a practical alternative. On the observational side, translating \((r_2,\lambda)\) into ringdown and tail shifts offers a path to constraints with \emph{LIGO–Virgo–KAGRA (LVK)} and \emph{LISA}, provided that targeted injection studies quantify detectability under realistic noise.

\section*{Acknowledgments}
This research was supported by the National Science, Research and Innovation Fund (NSRF) and Prince of Songkla University (Grant No. SCI6701314S). The author appreciates Takol Tangphati for introducing the asymptotic iteration method to the author's knowledge and H. Duerrast for proofreading.

\bibliographystyle{unsrturl}
\bibliography{references}

\end{document}